\documentclass[aps,preprint,nofootinbib,floatfix,superscriptaddress]{revtex4-2}
\usepackage{graphicx}
\usepackage[dvipsnames]{xcolor}
\usepackage[utf8]{inputenc}
\usepackage{amsmath,amssymb,mathtools}
\usepackage[pdfusetitle]{hyperref}
\usepackage{epstopdf}

\newcommand{\lsim}{\mathrel{\mathop{\kern 0pt \rlap
  {\raise.2ex\hbox{$<$}}}
  \lower.9ex\hbox{\kern-.190em $\sim$}}}
\newcommand{\gsim}{\mathrel{\mathop{\kern 0pt \rlap
  {\raise.2ex\hbox{$>$}}}
  \lower.9ex\hbox{\kern-.190em $\sim$}}}

\newcommand{\gev}{\ensuremath{\,\mathrm{GeV}}}
\newcommand{\tev}{\ensuremath{\,\mathrm{TeV}}}

\newcommand{\lavg}{\ensuremath{\left\langle}}
\newcommand{\ravg}{\ensuremath{\right\rangle}}

\newcommand*{\avg}[1]{\ensuremath{\lavg #1 \ravg}}

\begin{document}

\title{Relic density and temperature evolution of a light dark sector}
\author{Xin-Chen Duan}
\email{xcduan@pmo.ac.cn}
\affiliation{
Key Laboratory of Dark Matter and Space Astronomy,
Purple Mountain Observatory, Chinese Academy of Sciences, Nanjing 210008,
China
}
\affiliation{School of Astronomy and Space Science, University of Science and Technology of China, Hefei, Anhui 230026, China}

\author{Raymundo Ramos}
\email{raramos@kias.re.kr}
\affiliation{
Quantum Universe Center, Korea Institute for Advanced Study, Seoul 02455, Korea
}

\author{Yue-Lin Sming Tsai}
\email{smingtsai@pmo.ac.cn}
\affiliation{
Key Laboratory of Dark Matter and Space Astronomy,
Purple Mountain Observatory, Chinese Academy of Sciences, Nanjing 210008,
China
}
\affiliation{School of Astronomy and Space Science, University of Science and Technology of China, Hefei, Anhui 230026, China}

\date{\today}

\begin{abstract}
We have developed a set of four fully coupled Boltzmann equations to precisely determine the relic density and temperature of dark matter by including three distinct sectors: dark matter, light scalar, and standard model sectors.
The intricacies of heat transfer between dark matter (DM) and the standard model sector through a light scalar particle are explored, inspired by stringent experimental constraints on the scalar-Higgs mixing angle and the DM-scalar coupling. 
Three distinct sectors emerge prior to DM freeze-out, requiring fully coupled Boltzmann equations to accurately compute relic density. 
Investigation of forbidden, resonance, and secluded DM scenarios demonstrates significant deviations between established methods and the novel approach with fully coupled Boltzmann equations. Despite increased computational demands, this emphasizes the need for improved precision in relic density calculations, underlining the importance of incorporating these equations in comprehensive analyses.
\end{abstract}

\maketitle

\section{Introduction}

Exploring dark matter (DM) via interactions between the dark sector and the standard model (SM) sector aids in unraveling the nature of DM. 
Weakly interacting massive particles (WIMP), a potential DM candidate, are expected to produce a detectable signal in DM measurements; 
although, currently it remains elusive in recent findings from LHC~\cite{Felcini:2018osp}, XENON1T~\cite{Aprile:2018dbl}, and various DM indirect detection experiments~\cite{Ackermann:2015zua,Hoof:2018hyn,Oakes:2019ywx,Boudaud:2016mos,Roszkowski:2017nbc}. 
Among these experimental attempts, the measurement of DM relic density from cosmic microwave background (CMB) radiation stands out as it provides both upper and lower limits on the DM interaction rate. 
In other words, the DM annihilation rate is required to fall within a specific range to match the relic density reported by PLANCK~\cite{Aghanim:2018eyx}. 
Considering this, by assuming a certain coupling strength for the DM-SM interaction, the DM mass can be constrained to belong in a particular range. 
For instance, a DM particle with weak coupling typically has a mass around $100\gev$, known as the WIMP miracle~\cite{Feng:2008ya}, 
while strong coupling is associated with a DM mass around $200\tev$~\cite{Huo:2015nwa}. 
Therefore, the application of the relic density constraint significantly narrows down the allowed parameter space for DM models, facilitating the search for DM signals.

The calculation of relic density is heavily dependent on the thermal history of the dark sector. 
The most straightforward assumption is the thermal dark matter paradigm, where the observed relic abundance can be naturally explained by the freeze-out mechanism. 
During the radiation-dominated era, thermal DM is generated through collisions within the thermal plasma, with its number density following the thermal Boltzmann distribution. 
A similar successful explanation can be applied to the history of Big Bang Nucleosynthesis if SM particles adhere to the same assumption. 
However, recent XENON1T findings impose a stringent limit on the DM-nucleon cross-section, rendering some parameter space with the interaction too tiny to maintain thermal equilibrium prior to freeze-out, a phenomenon known as ``early kinetic decoupling''.
In this regard,
early kinetic decoupling is investigated, focusing on DM resonant annihilation~\cite{Binder:2017rgn,Duch:2017nbe,Brahma:2023psr,Ghosh:2023mnh} and forbidden DM annihilation~\cite{Liu:2023kat,Aboubrahim:2023yag}. 
Studies on thermal freeze-out mechanisms involving vector DM, such as those on the $U(1)_{X}$ Higgs portal~\cite{Amiri:2022cbv} and vector-portal frameworks~\cite{Fitzpatrick:2020vba}, relax the assumption of thermal equilibrium between the dark and SM sectors during DM freeze-out. 
Furthermore, researchers have investigated the impact of early kinetic decoupling in a fermionic dark matter model with CP-conserving and CP-violating interactions mediated by Higgs exchange~\cite{Abe:2020obo} as well as in a pseudo-Nambu-Goldstone model~\cite{Abe:2021jcz}.

The standard calculation for relic density assumes that the DM initial density is in thermal equilibrium at a temperature slightly above the freeze-out temperature. 
This assumption is utilized in software packages like \texttt{DarkSUSY}~\cite{Bringmann:2022vra}, \texttt{micrOMEGAs}~\cite{Belanger:2018ccd}, and \texttt{MadDM}~\cite{Ambrogi:2018jqj}. 
However, this approach neglects the effects of early kinetic decoupling and temperature differences between the dark and SM sectors, as demonstrated in Ref.~\cite{Binder:2017rgn}. 
To accurately calculate relic density for DM with early kinetic decoupling, one must consider the Boltzmann equations for the density and temperature of the dark sector and initiate the evolution from a temperature significantly higher than the freeze-out temperature.

To produce the correct relic density with DM mass lighter than the Lee-Weinberg bound~\cite{Lee:1977ua}, 
one can consider
a simplified DM model that contains a DM candidate and a light mediator. 
In a simplified Higgs portal model~\cite{Matsumoto:2018acr, Chen:2024njd},
the DM candidate, $\chi$, is a light Majorana particle
and the light mediator, $\phi$, is a scalar, both of them singlet under the SM.
Due to mixing between the light scalar and the SM Higgs boson
the DM candidate can interact with the SM mediated by the light scalar singlet $\phi$.
The mixing between the SM Higgs boson and $\phi$ is characterized by a mixing angle, $\theta$. 
These two new particles are expected to be produced in meson decay processes 
at LHCb~\cite{LHCb:2012juf,LHCb:2016awg} and beam dump experiments~\cite{KTEV:2000ngj, KTeV:2003sls, Bezrukov:2009yw, BNL-E949:2009dza, BaBar:2015jvu, KOTO:2018dsc, NA62:2020xlg, Belle-II:2023esi}. 
However, none of them has been observed putting a severe exclusion on the model parameter space,
as demonstrated
by a comprehensive likelihood analysis involving a robust set of constraints~\cite{Matsumoto:2018acr, Chen:2024njd}. 
Consequently, a large portion of the surviving parameter space can have early kinetic decoupling between the dark, scalar, and SM sectors because of 
the suppressed $\sin\theta$. 
In such parameter space, it would be interesting to perform a precise study of early kinetic decoupling.

Based on the model in Ref.~\cite{Matsumoto:2018acr}, we will revise the relic density computation 
by allowing different temperature evolution for the dark, $\phi$, and SM sectors. 
We focus on three typical scenarios where the leading DM annihilation in the early universe are: 
\begin{itemize}
    \item [(i)] forbidden annihilations $\chi+\chi\to \phi+\phi$ with subsequent decays mainly to SM states 
where DM is only slightly lighter than $\phi$~\cite{DAgnolo:2015ujb} 
accompanied by a relatively large mixing angle $\sin\theta\thickapprox\mathcal{O}(10^{-3})$,
    \item [(ii)] resonance annihilation scenario $\chi+\chi\to\phi\to {\rm SM+SM}$ enhanced at a certain DM temperature via 
$\phi$ resonance with $2 m_\chi\thickapprox m_\phi$,
    \item [(iii)] secluded DM scenario~\cite{Pospelov:2007mp} 
    involving a tiny mixing angle and $m_{\phi}\ll m_{\chi}$. 
Importantly, a considerable interaction between the $\phi$ and dark sector in this scenario is necessary to maintain 
thermal equilibrium before the DM chemical decoupling.
\end{itemize}
Although we restrict ourselves to the light scalar, our derivation of Boltzmann equations is 
generic and can be applied to the light vector mediator as well.

This paper is structured as follows. 
In Sec.~\ref{sec:model}, we provide a brief overview of the Majorana DM model incorporating a light singlet scalar and highlight its main features. 
In Sec.~\ref{sec:Boltzmann}, we adapt the standard Boltzmann equation to incorporate the evolution equations for the number density and temperature of both DM and the new scalar. 
Sec.~\ref{sec:result} presents our numerical results for three selected benchmark scenarios. 
Finally, Sec.~\ref{sec:conclusion} summarizes and discusses our findings. 
Additionally, we include detailed derivations of the Boltzmann equations in the appendices at the end of this paper.
\section{Majorana dark matter with a light singlet scalar mediator}
\label{sec:model}

In this section, we review the minimal Higgs portal Model described in Refs.~\cite{Matsumoto:2018acr, Chen:2024njd}, which considers a light fermionic WIMP, $\chi$, with a mass around $\mathcal{O}(1)$ GeV or less. 
This Majorana WIMP requires the presence of a light mediator, $\Phi$, chosen as a real singlet scalar for simplicity. 
Both the Majorana DM field and the singlet scalar mediator are singlets under the SM gauge group. 
Additionally, the Majorana DM is associated with the odd sector of an imposed $\mathbb{Z}_2$ symmetry, 
while all other particles are even under this symmetry. 
The minimal renormalizable Lagrangian, which includes a light Majorana DM, a real scalar mediator, and all SM interactions, is expressed as 
\begin{equation}
	\mathcal{L} =
	\mathcal{L}_{\rm SM} + \frac{1}{2} \bar{\chi} (i\gamma_\mu\partial^\mu - m_{\chi}) \chi + 
	\frac{1}{2} (\partial \Phi)^2
	- \frac{c_s}{2} \Phi \bar{\chi} \chi - \frac{c_p}{2} \Phi \bar{\chi} i \gamma_5 \chi
	-V(\Phi,H).
	\label{eq:lagrangian}
\end{equation}
The SM lagrangian is represented by $\mathcal{L}_{\rm SM}$ and $H$ is the SM
Higgs doublet.
The full scalar potential is given by $V(\Phi,H) \equiv V_\Phi(\Phi) + V_{\Phi
H}(\Phi, H)$ plus the potential of $H$, $V_H(H)$, contained in $\mathcal{L}_{\rm
SM}$.
These components of the potential are given by
\begin{align}
	V_H(H) &= \mu^2_H H^{\dagger} H + \frac{\lambda_H}{2} (H^{\dagger} H)^2,
	\nonumber \\
	V_{\Phi}(\Phi) &= \mu^3_1 \Phi + \frac{\mu^2_\Phi}{2} \Phi^2 + \frac{\mu_3}{3!} \Phi^3 + \frac{\lambda_\Phi}{4!} \Phi^4,
	\nonumber \\
	V_{\Phi H}(\Phi, H) &= A_{\Phi H} \Phi H^{\dagger} H + \frac{\lambda_{\Phi H}}{2} \Phi^2 H^{\dagger} H,
	\label{eq:potential}
\end{align}
where $\lambda_{H,\Phi,\Phi H}$ are dimensionless couplings, but $\mu_{H,1,\Phi,3}$ and $A_{\Phi H}$ couplings have mass dimension one.
Following the procedure of Ref.~\cite{Matsumoto:2018acr, Chen:2024njd}, we also assume
that the vacuum expectation value (VEV) of $\Phi$, $\langle\Phi\rangle =
v_\Phi$, vanishes such that $\Phi = v_\Phi + \phi' = \phi'$.
Furthermore, we use the unitary gauge expansion of the Higgs doublet
around its VEV, $H = [0, (v_H + h')/\sqrt{2}]^T$.
Under these assumptions, the minimization conditions for the potential result
in the following expressions for $\mu^2_H$ and $\mu^3_1$ as functions of other
parameters
\begin{align}
\label{eq:minmuH2}
\frac{\lambda_H v_H^2}{2} + \mu_H^{2} = 0 \:\:\:\:& \Rightarrow \:\:\:\:
\mu^2_H = - \frac{\lambda_H v_H^2}{2},\\
\label{eq:minmu13}
\frac{A_{\Phi H} v_H^2}{2} + \mu_1^3  = 0 \:\:\:\:& \Rightarrow \:\:\:\:
\mu^3_1 = - \frac{A_{\Phi H} v_H^2}{2}.
\end{align}

From the quadratic terms in the scalar potential we can find the following
squared mass matrix written in the $\{h', \phi'\}$ basis
\begin{equation}
\label{eq:sqrdmass}
\left(
	\begin{matrix}
		\lambda_{H} v_{H}^{2} & A_{\Phi H} v_{H} \\
		A_{\Phi H} v_{H} & \frac{\lambda_{\Phi H} v_{H}^{2}}{2} + \mu_{\Phi}^{2}
	\end{matrix}
\right) =
U^T_\theta
\left(
	\begin{matrix}
		m^2_h & 0 \\
		0 & m^2_\phi
	\end{matrix}
\right)
U_\theta,\:\:\:\:
\text{with}\:\:\:\:
U_\theta = 
\left(
	\begin{matrix}
		\cos\theta & - \sin\theta \\
		\sin\theta & \cos\theta
	\end{matrix}
\right),
\end{equation}
where $U_\theta$ is the matrix that relates the interaction states $h'$ and
$\phi'$ with the mass eigenstates $h$ and $\phi$: $(h, \phi)^T = U_\theta (h',
\phi')^T$.
The mixing angle $\theta$ and the eigenvalues of the squared mass matrix are
given by
\begin{align}
\tan 2\theta & = \frac{4 A_{\Phi H} v_H}{\lambda_{\Phi H}
v_H^2 + 2 \mu_\Phi^2 - 2 \lambda_H v_H^2 }\\
m^2_{h,\phi} & = \frac{1}{2}\left\{
	v_H^2\left(\lambda_H + \frac{\lambda_{\Phi H}}{2}\right) + \mu_\Phi^2
	\pm \sqrt{
		\left[v_H\left(\lambda_H - \frac{\lambda_{\Phi H}}{2}\right)
		- \mu_\Phi^2\right]^2 + 4A_{\Phi H}^2 v_H^2
	}
\right\}.
\end{align}
By comparing the left-hand-side and right-hand-side of Eq.~\eqref{eq:sqrdmass}
we can trade three model parameters with the mass eigenvalues $m_h^2$ and
$m_\phi^2$ and the mixing angle $\theta$.
Namely, the three model parameters $\lambda_H$, $A_{\Phi H}$ and
$\lambda_{\Phi H}$ can be expressed as
\begin{alignat}{2}
\lambda_H & = \frac{m_H^2 +  \left(m_\phi^2 - m_H^2\right)
	\sin^2\theta}{v_H^2}
	&& 
 \:\:\:\: \stackrel{\theta\approx 0}{\Rightarrow} \:\:\:\: 
 \frac{m_H^2}{v_H^2}, \\
	A_{\Phi H} & = \frac{\left(m_\phi^2 - m_H^2\right)
	\sin\theta\cos\theta}{v_H} 
 && 
\:\:\:\:\stackrel{\theta\approx 0}{\Rightarrow} \:\:\:\: 
0,\\
\lambda_{\Phi H} & = \frac{2 \left[m_\phi^2 - \mu_\Phi^2
		- \left(m_\phi^2 - m_H^2\right) \sin^2\theta\right]}{v_H^2}
	&& 
\:\:\:\:\stackrel{\theta\approx 0}{\Rightarrow} \:\:\:\:
 \frac{2 \left[m_\phi^2 - \mu_\Phi^2 \right]}{v_H^2},
\end{alignat}
where the rightmost expressions represent the $\theta \to 0$ limit (the SM limit).

Note that by taking the mass of the Higgs to be its measured value,
$m_h\approx 125$~GeV, and the SM VEV fixed at $v_H = 246$~GeV, the parameter
$\lambda_H$ is fixed in the $\sin^2\theta \to 0$ limit.
Thus, we are left with five free parameters from the scalar potential:
$\theta$, $m_\phi$, $\mu_\phi^2$, $\mu_3$ and $\lambda_\Phi$.
To those parameters, we add the WIMP mass $m_\chi$ and the couplings
between the WIMP and the scalar $\Phi$, $c_s$ and $c_p$, leaving a total of
eight free parameters
\begin{equation}
\lambda_{\Phi H}, A_{\Phi H}, \mu_\phi^2, \mu_3, \lambda_\Phi, m_\chi, c_s, c_p
\:\:\:\:\:\to\:\:\:\:\:
\theta, m_\phi, \lambda_{\Phi H}, \mu_3, \lambda_\Phi, m_\chi, c_s, c_p
\end{equation}

The interactions between the scalars can be extracted from the Lagrangian of
Eq.~\eqref{eq:lagrangian}.
After moving to the mass eigenstates base and using $m_\phi$ and $\theta$ as
free variables we can rewrite the three-scalars terms in the Lagrangian as
\begin{equation}
-\frac{c_{hhh}}{3!}h^3
-\frac{c_{\phi hh}}{2}\phi h^2
-\frac{c_{\phi \phi h}}{2}\phi^2 h
-\frac{c_{\phi \phi \phi }}{3!}\phi^3
\end{equation}
where the couplings are given
\begin{align}
\label{eq:chhh}
c_{hhh} & = v_H^{-1}c_\theta\left[
	3 m_H^2 \left(c_\theta^2 + 2 s_\theta^{4}\right)
	+ 6 m_\phi^2 s_\theta^2 c_\theta^2
	- 6 \mu_{\Phi}^2 s_\theta^2
\right] - \mu_3 s_\theta^3, \\
\label{eq:cphh}
c_{\phi h h} & = v_H^{-1}s_\theta \left[
	2 m_H^2 (1 - 3 s_\theta^2c_\theta^2)
	- 3 m_\phi^2 c_\theta^2 (1 - 2 s_\theta^2)
	+ 2 \mu_{\Phi}^2(2 - 3 s_\theta^2)
\right] + \mu_3 c_\theta s_\theta^2, \\
\label{eq:cpph}
c_{\phi\phi h} & = v_H^{-1} c_\theta\left[
	3 m_H^2 s_\theta^2 (1 - 2 s_\theta^2)
	+ 2 m_\phi^2 (1 - 3 s_\theta^2 c_\theta^2)
	- 2 \mu_\Phi^2 (1 - 3 s_\theta^2)
\right] - \mu_3 s_\theta c_\theta^2, \\
\label{eq:cppp}
c_{\phi\phi\phi} & = v_H^{-1}s_\theta \left[
	6 m_H^2 s_\theta^{2} c_\theta^2
	+ 3 m_\phi^2 (2 + 2 s_\theta^{4} - 3 s_\theta^{2})
	- 6 \mu_{\Phi}^{2} c_\theta^{2}
\right] + \mu_3 c_\theta^3.
\end{align}
It is straightforward to write the $s_\theta\equiv\sin\theta\to 0$ and $c_\theta\equiv\cos\theta\to 1$ limits of these expressions:
\begin{align}
\label{eq:chhh_s2t0}
c_{hhh} & \sim 3 v_H^{-1} c_\theta m_H^2\\
\label{eq:cphh_s2t0}
c_{\phi h h} & \sim v_H^{-1} s_\theta (2 m_H^2 - 3 m_\phi^2 + 4 \mu_{\Phi}^2) \\
\label{eq:cpph_s2t0}
c_{\phi\phi h} & \sim 2 v_H^{-1} c_\theta\left(m_\phi^2 - \mu_{\Phi}^2\right) - \mu_3 s_\theta\\
\label{eq:cppp_s2t0}
c_{\phi\phi\phi} & \sim 6 v_H^{-1} s_\theta \left( m_\phi^2 -  \mu_{\Phi}^{2} \right) + \mu_{3} c_\theta
\end{align}

The couplings of Eqs.~\eqref{eq:chhh}--\eqref{eq:cppp} are equivalent to the
expressions found below Eq.~(2.7) of Ref.~\cite{Matsumoto:2018acr} rewritten
to use $\theta$, $m_H$ and $m_\phi$ instead of $\lambda_H$, $A_{\Phi H}$ and
$\lambda_{\Phi H}$.
The four-point interactions are only relevant for very massive $\chi$ and can
be consulted in and below Eq.~(2.8) of Ref.~\cite{Matsumoto:2018acr}.
Finally, due to the mixing between $h'$ and $\phi'$, the terms of the
Lagrangian that couple the DM $\chi$ and the scalar mass eigenstates are given
by
\begin{equation}
\mathcal{L}_\text{int} \supset -\frac{1}{2}\left[
	c_\theta \phi \left(c_s \bar{\chi} \chi
		+ i c_p \bar{\chi} \gamma_5 \chi\right)
	+ s_\theta h \left(c_s \bar{\chi} \chi
		+ i c_p \bar{\chi} \gamma_5 \chi\right)
\right].
\end{equation}
It is easy to see that as long as we keep $s_\theta$ as a small parameter, the
coupling between the Higgs $h$ and $\chi$ remains suppresed.
As it was mentioned in Ref.~\cite{Matsumoto:2018acr}, precision measurements
on the properties of the Higgs found at the LHC require $|\theta|\ll 1$.

\section{Boltzmann equations}
\label{sec:Boltzmann}

For a particle species $i$, the evolution of its phase space density $f_i(t, \mathbf{p})$ 
is governed by the Boltzmann equation~\cite{Bringmann:2006mu},
which, in a Friedmann-Robertson-Walker universe, takes the form of
\begin{equation}
    \label{eq:boltzmann}
    E \left(
        \partial_t - H \mathbf{p} \cdot \nabla_\mathbf{p}
        \right) f_i
    = C_i[f_i].
\end{equation}
Here, $H \equiv (1/a) (da/dt)$ is the Hubble parameter, $a(t)$ is the scale factor of the universe, and $t$ is the evolution time. 
The $i$ particle possesses energy $E = E_i$ and three-momentum $\mathbf{p} = \mathbf{p}_i$. On the right side, the collision term $C_i[f_i]$ contains all interactions of particle $i$. Our focus here is solely on the Boltzmann equations for two weakly interacting particles, $\chi$ and $\phi$. 
The dominant processes in the collision terms include annihilation, elastic scattering, $\phi$ decay, and $\phi$ absorption. By following the approach in Ref.~\cite{Binder:2017rgn}, we can derive complete collision terms, as shown in Appendix~\ref{sec:coll}, and evolution equations as shown in Appendix~\ref{sec:0th_and_2nd}.

\begin{figure}
\begin{center}
\includegraphics[width=1.0\textwidth]{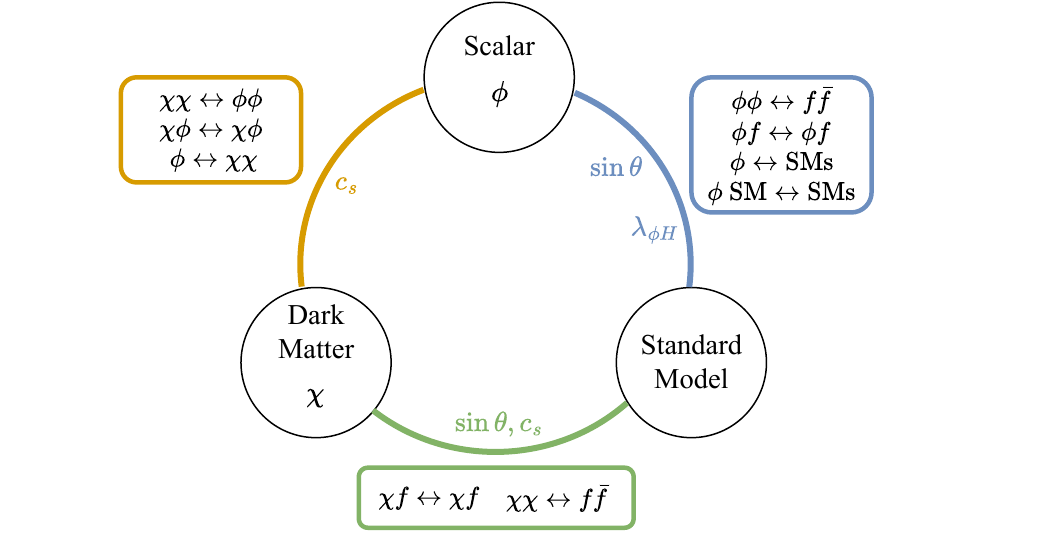}
\caption{A schematic plot to illustrate the interactions between three sectors. 
The parameters shown near the color lines govern the relevant processes in the nearest box.
\label{Fig:schematic}}
\end{center}
\end{figure} 

Before delving into our primary equations, it is crucial to highlight the difference between our computational approach and conventional methods like \texttt{DarkSUSY}~\cite{Bringmann:2018lay} and \texttt{micrOMEGAs}~\cite{Belanger:2018mqt} for calculating relic density. 
These programs assume that $\phi$ interacts sufficiently with both the SM and dark sectors to maintain thermal equilibrium during freeze-out. 
However, our approach differs as the thermal history of $\phi$ and $\chi$ evolves independently from the SM sector. 
Practically, this implies solving a temperature equation alongside the number density equation for each particle, resulting in four coupled Boltzmann equations governing the evolution of $\chi$ and $\phi$.

Throughout our work, we denote the temperatures of the SM, $\phi$, and dark sectors as $T$, $T_\phi$, and $T_\chi$ respectively.~\footnote{Since the SM sector is fully in thermal equilibrium before DM freeze-out, the SM temperature $T$ is identical to the photon temperature.}
If the three sectors are in thermal equilibrium ($T = T_\chi = T_\phi$), one can simply evolve only the number densities of $\chi$ and $\phi$ with respect to $T$ near the freeze-out temperature $T_F$. 
However, in our case, we have to consider three interaction rates among three sectors: $\Gamma(\phi\leftrightarrow\chi)$, $\Gamma(\phi\leftrightarrow\mathrm{SM})$, and $\Gamma(\chi\leftrightarrow\mathrm{SM})$, as illustrated in Fig.~\ref{Fig:schematic}. 
If $\Gamma(\chi\leftrightarrow\mathrm{SM})$ or 
$\min\left[\Gamma(\phi\leftrightarrow\chi), \Gamma(\phi\leftrightarrow\mathrm{SM})\right]$ is not greater than the Hubble expansion rate, then thermal equilibrium between the DM and the SM sectors cannot be maintained, leading to kinetic decoupling of the DM from the SM sector.

In such case, the standard relic density computation, based solely on DM number density evolution, can be inaccurate.

In standard calculations, the initial number density condition for a particle with mass $m_i$ relies on the thermal distribution $\exp\left(-m_i/T\right)$ at freeze-out. Nonetheless, this approach becomes inadequate when early kinetic decoupling is considered. Hence, it is necessary to revert to the original assumption for thermal DM, where particles are presumed to be in equilibrium with the SM sector as early as $T\approx m_i$.
The standard form of the distribution at thermal equilibrium (maximum entropy) is 
\begin{equation}
    f(E_i,T)=\left[\exp\left(\frac{E_i-\mu}{T}\right) \pm 1\right]^{-1}, 
    \label{eq:fE}
\end{equation}
where the plus sign corresponds to fermions, and the minus sign corresponds to bosons.
These statistical mechanical factors can be safely neglected as the particles are non-relativistic within the subsequently considered range. 
In the secluded scenario, despite $\phi$ being relativistic, two conditions ensure that $E_\phi$ remains much greater than $T$: energy conservation in the process $\phi \phi \leftrightarrow \chi \chi$ and thermal equilibrium between $\chi$ and $\phi$. 
This also allows the statistical mechanical factor to be neglected.

The number density $n_i$ and temperature $T_i$ of a particle $i$ are defined as

\begin{equation}
\label{eq:n_and_T}
	n_i \equiv
	g_i \int \frac {d^3 \mathbf{p}_i} {(2\pi)^3}
	f_i (E_i, T), 
	\quad\text{and}\quad
	T_i
	\equiv
	\avg{
		\frac {\mathbf{p}_i^2 } {3 E_i}
	}
	=
	\frac {g_i} {n_i}
	\int \frac {d^3 \mathbf{p}_i} {(2\pi)^3}
	\frac {\mathbf{p}_i^2 } {3 E_i}
	f_i (E_i, T), 
\end{equation}
respectively.
In Eq.~\eqref{eq:n_and_T} above,
the dependence on momentum is left implicit in $E_i = E_i(\mathbf{p}_i)$.
For simplicity, we can represent $f_i(E_i, T)$ as $f(E_i)$ when $T$ matches the temperature of particle $i$. 
Before and during chemical freeze-out, we express the phase space distribution $f(E_i, T_i)$ in terms of $f_{\mathrm{eq}}(E_i, T_i)$ using the equation: 
\begin{equation}
\label{eq:ansatz}
    f(E_i,T_i) = \frac{n(T_i)} {n_{\mathrm{eq}}(T_i)} f_{\mathrm{eq}}(E_i,T_i).
\end{equation}
When $T_i=T$ and $n(T_i)/n_{\mathrm{eq}}(T)\approx 1$, full equilibrium between the SM sector and the $i$ sector is implied.

Finally, we introduce two dimensionless quantities
\begin{equation}
	Y_i \equiv \frac {n_i} {s} \quad
	\textrm{and} \quad
	y_i \equiv \frac {m_i} {s^{2/3}} T_i
	\label{eq:Yy}
\end{equation}
corresponding to the comoving number density and the comoving temperature. 
Here, $s=(2\pi^2/45) g_\mathrm{eff}^s T^3$ is the entropy density of the universe with
effective entropy degrees of freedom $g_\mathrm{eff}^s$.
The evolution equations for $Y_i$ and $y_i$ can be obtained from the Boltzmann Eq.~\eqref{eq:boltzmann}. 
Taking $i$ equal to $\chi$ or $\phi$, we can express the comoving number density and temperature equations in terms of annihilation cross sections and decay widths, as explained in detail in Appendix~\ref{sec:0th_and_2nd}.

\subsection{The evolution equations of the comoving number density}
\label{sec:BE_n}

The temperature of DM, SM, and $\phi$ can be all different and we represent their temperatures as $T_\chi$, $T$, 
and $T_\phi$, respectively. 
By introducing the derivative $Y_i'=dY_i/dx$ 
with respect to the dimensionless variable $x \equiv m_\chi/T$, the comoving number density evolution equations for $\chi$ and $\phi$ are given by 
\begin{align}
\label{numberBEs}
	x \tilde{H} 
	Y_\chi^{'}
	=& \avg{
		\sigma_{\varphi \Bar{\varphi}\to \chi \chi } v
	}_T
	s Y_{\varphi, \mathrm{eq}}^2
	- \avg{
		\sigma_{\chi \chi \to \varphi \Bar{\varphi}} v
	}_{T_\chi}
	s Y_{\chi}^2 
\nonumber\\
- &  \avg{\sigma_{\chi \chi \to \phi \phi} v }_{T_\chi}
	s Y_{\chi}^2 +
	\avg{\sigma_{\phi \phi \to \chi \chi} v}_{T_\phi}
	s Y_\phi^2  
\nonumber\\
 	+ & 
 \avg{
		\Gamma_{\phi \to \chi \chi}
	}_{T_\phi}
	Y_{\phi}
	- 
	\avg{
	\sigma_{\chi \chi \to \phi} v
	}_{T_\chi} s Y_{\chi}^2
\end{align}
and
\begin{align}
\label{numberBEs2}
    x \tilde{H} 
	Y_\phi^{'}
	=& \avg{
		\sigma_{\varphi \Bar{\varphi}\to \phi \phi} v
	}_{T}
	s Y_{\varphi, \mathrm{eq}}^2
	- \avg{
		\sigma_{\phi \phi \to \varphi \Bar{\varphi}} v
	}_{T_\phi}
	s Y_{\phi}^2
	\nonumber\\
	- & \avg{
		\sigma_{\phi \phi \to \chi \chi} v
	}_{T_\phi}
	s Y_{\phi}^2
	+ \avg{
			\sigma_{\chi \chi \to \phi \phi} v
		}_{T_\chi}
	s Y_\chi^2
	\nonumber\\
	- & \avg{
		\Gamma_{\phi \to \varphi\bar{\varphi} }
	}_{T_\phi}
	Y_{\phi}
	+ 
\avg{
		\sigma_{\varphi \Bar{\varphi}\to \phi} v
	}_{T}
	s Y_{\varphi, \mathrm{eq}}^2
	\nonumber\\
    - &  \avg{
		\Gamma_{\phi \to \chi \chi}
	}_{T_\phi}
	Y_{\phi}
	+ 
	\avg{
	\sigma_{\chi \chi \to \phi} v
	}_{T_\chi} s Y_{\chi}^2
    \nonumber\\
    + &
    \sum_{\varphi_2,\varphi_3,\varphi_4}
    \left[
        \avg{
                \sigma_{\varphi_3 \varphi_4 \to \phi \varphi_2 } v
        }_T\ 
        s Y_{\varphi_3, \mathrm{eq}} Y_{\varphi_4, \mathrm{eq}}
    -  \avg{
                \sigma_{\phi \varphi_2 \to \varphi_3 \varphi_4} v
        }_{(T_\phi,T)}\ 
    s Y_{\varphi_2, \mathrm{eq}} Y_\phi \right].
\end{align}
The Hubble parameters used in this work are 
\begin{equation}
	\tilde{H} \equiv 
	H \left[1+ \frac{1}{3}
	\frac { d\log(g_{\mathrm{eff}}^s) } {d\log(T)} \right]^{-1},\quad {\rm and}\quad
 H = \sqrt{\frac{4\pi^{3}g_{\mathrm{eff}}^s}{45}}\frac{T^{2}}{m_{\rm planck}}, 
\end{equation}
where $m_{\rm planck}$ is the Planck mass.
The particles ($\varphi_i$s) involved in the co-annihilation processes are SM particles, including gluons $g$, SM fermions $f$ like leptons $l^\pm$ and quarks $q$, photons $\gamma$, and Higgs boson.
The co-annihilation contributions can be dominated by the process $q+g \leftrightarrow f+\phi$ before the QCD epoch 
and $f+\gamma\leftrightarrow f+\phi$ after the QCD epoch.

We define the thermally averaged annihilation cross-section with temperature $T_i$ and $T_j$ as 
\begin{align}
\label{eq:sigmav0}
\avg{\sigma_{i j \to a b} v}_{(T_i,T_j)} \equiv
	\frac {g_i} {n_{i, \mathrm{eq}} (T_i)}
	\frac {g_j} {n_{j, \mathrm{eq}} (T_j)}
	\int &
	\frac {d^3\mathbf{p}_i} {(2\pi)^3}
	\frac {d^3\mathbf{p}_j} {(2\pi)^3}
	\sigma_{ij\to ab}\, v_\text{M\o l}\,
	f_{\mathrm{eq}}(E_i,T_i) f_{\mathrm{eq}}(E_j,T_j)
	\nonumber\\
	=
	\frac {1} {n_{i, \mathrm{eq}} (T_i)}
	\frac {1} {n_{j, \mathrm{eq}} (T_j)}
	\int &
	d\Pi_{i}d\Pi_{j}d\Pi_{a}d\Pi_{b}
	(2 \pi)^{4} \delta^{(4)} (p_i + \tilde{p}_j - k_a - \tilde{k}_b)
	\nonumber\\
	&
	\times |\mathcal{M}_{ij \to ab}|^2
	f_{\mathrm{eq}}(E_i,T_i) f_{\mathrm{eq}}(E_j,T_j)\\
	=
	\frac {1} {n_{i, \text{eq}} (T_i)}
	\frac {1} {n_{j, \text{eq}} (T_j)}
	\int
	&
	\frac{dE_+ dE_- ds\, dc_\theta^*}{128\pi (2\pi)^4 s}
	\sqrt{\lambda_{ab}} |\mathcal{M}_{ij \to ab}|^2
	f_{\text{eq}}(E_i,T_i) f_{\text{eq}}(E_j,T_j),
	\label{eq:sigmav3}
\end{align}
where the phase space volume element and the number density at thermal equilibrium for each particle are   
\begin{equation}
\label{eq:dPi_BEq}
d\Pi_i\equiv \frac {d^3 \mathbf{p}_i} { (2\pi)^3 2 E_i},\quad {\rm and}\quad 
n_{i, \mathrm{eq}} \equiv 
	g_i \int \frac {d^3 \mathbf{p}_i} {(2\pi)^3}
    f_{i,\mathrm{eq}}(E_i).
\end{equation}
The index $i,j$ are the initial state particles while $a,b$ are final state particles. 
If temperature $T_i$ and $T_j$ are the same, we denote $\avg{\sigma v}_T\equiv\avg{\sigma v}_{(T,T)}$. 
The integration in Eq.~\eqref{eq:sigmav3} is performed for squared center-of-mass energy $s= (p_i+p_j)^2 \geq \max[(m_i + m_j)^2, (m_a + m_b)^2]$,
$E_+\equiv E_i + E_j \geq \sqrt{s}$, $-1 < c_\theta^* < 1$ and 
\begin{equation}
\frac{E_+(m_j^2 - m_i^2) - \sqrt{\lambda_{ij}}\sqrt{E_+^2 - s}}{s} \leq
E_- \leq
\frac{E_+(m_j^2 - m_i^2) + \sqrt{\lambda_{ij}}\sqrt{E_+^2 - s}}{s}.
\end{equation}
where $E_{-} \equiv E_i - E_j$.  
The cross-section is then given by  
\begin{equation}
\label{eq:xsec}
\sigma_{ij\to ab} \equiv \frac{1}{2 g_i g_j\sqrt{\lambda_{ij}}}
    \int |\mathcal{M}_{ij \to ab}|^2
    (2 \pi)^{4} \delta^{(4)} (p_i + \tilde{p}_j - k_a - \tilde{k}_b) 
    d\Pi_{a}d\Pi_{b}  
\end{equation}
with Kallen function
\begin{equation}
\lambda_{ij}\equiv (s-m_i^2-m_j^2)^2-4 m^2_i
m^2_j = 4 v^2_{\text{M\o l}} E^2_i E^2_j.
\label{eq:Kallen}
\end{equation}

\subsection{The evolution equations of the temperature}
\label{sec:BE_t}
Next, the temperature evolution is governed by the equations
\begin{align}
\label{tempBEs}
	x \tilde{H} Y_\chi T_\chi
	\left(
		\frac{y_\chi^{'}} {y_\chi}
		+ \frac{Y_\chi^{'}} {Y_\chi}
	\right)
	= {}&\frac{H} {3}
	\avg{
		\frac {\mathbf{p}_\chi^4} {E_\chi^3}
	} Y_\chi +
\avg{
	T_\chi\sigma_{\varphi \Bar{\varphi}\to \chi \chi } v
	}_T
	s Y_{\varphi, \mathrm{eq}}^2
	- \avg{T_\chi
		\sigma_{\chi \chi \to \varphi \Bar{\varphi}} v
	}_{T_\chi}
	s Y_{\chi}^2 
\nonumber\\
- &  \avg{T_\chi\sigma_{\chi \chi \to \phi \phi} v }_{T_\chi}
	s Y_{\chi}^2 +
	\avg{T_\chi\sigma_{\phi \phi \to \chi \chi} v}_{T_\phi}
	s Y_\phi^2  
\nonumber\\
 	+ & \avg{
		T_\chi\Gamma_{\phi \to \chi \chi}
	}_{T_\phi}
	Y_{\phi}
	- 
	\avg{
	T_\chi\sigma_{\chi \chi \to \phi} v
	}_{T_\chi} s Y_{\chi}^2	
	\nonumber\\
	+ & \mathcal{S}_{\chi\phi}(T_\chi,T_\phi) sY_\chi Y_\phi + 
\mathcal{S}_{\chi\varphi}(T_\chi,T) s Y_\chi Y_{\varphi,\mathrm{eq}}\,
\end{align}
and
\begin{align}
\label{tempBEs2}
    x \tilde{H} Y_\phi T_\phi
	\left(
		\frac{y_\phi^{'}} {y_\phi}
		+ \frac{Y_\phi^{'}} {Y_\phi}
	\right)
	= {}&
	\frac{H}{3}
	\avg{
		\frac {\mathbf{p}_\phi^4} {E_\phi^3}
	} Y_\phi
	+
    \avg{
	T_\phi\sigma_{\varphi \Bar{\varphi}\to \phi \phi } v
	}_T
	s Y_{\varphi, \mathrm{eq}}^2
	- \avg{T_\phi
		\sigma_{\phi \phi \to \varphi \Bar{\varphi}} v
	}_{T_\phi} s Y_{\phi}^2 
	\nonumber\\
   - & \avg{T_\phi\sigma_{\phi \phi \to \chi \chi} v }_{T_\phi} s Y_{\phi}^2 
   + \avg{T_\phi\sigma_{\chi \chi \to \phi \phi} v}_{T_\chi} s Y_\chi^2  
\nonumber\\
 	- & \avg{ T_\phi\Gamma_{\phi} }_{T_\phi} Y_{\phi}
	+   \avg{ T_\phi\sigma_{\chi \chi \to \phi} v }_{T_\chi} s Y_{\chi}^2	
    +   \avg{ T_\phi\sigma_{\varphi \bar{\varphi} \to \phi} v }_{T} s Y_{\varphi,\mathrm{eq}}^2	
\nonumber\\
	+ & \mathcal{S}_{\phi\chi}(T_\phi,T_\chi) s Y_\chi Y_\phi + 
	    \mathcal{S}_{\phi\varphi}(T_\phi,T) s Y_\phi Y_{\varphi,\mathrm{eq}}
	\nonumber\\
	+ &
\sum_{\varphi_2,\varphi_3,\varphi_4}
    s\left[
        \avg{T_\phi\sigma_{\varphi_3 \varphi_4 \to \phi \varphi_2 } v
        }_T\ 
        Y_{\varphi_3, \mathrm{eq}} Y_{\varphi_4, \mathrm{eq}}
    -  \avg{T_\phi\sigma_{\phi \varphi_2 \to \varphi_3 \varphi_4} v
        }_{(T_\phi,T)}\ 
    Y_{\varphi_2, \mathrm{eq}} Y_\phi \right].
\end{align}
Considering that the temperature of the sector, e.g., $T_\chi$ for the dark sector, is included in the integral of the thermal average,   
we define a new physical quantity $T_\chi \sigma v$ for convenience as   
\begin{align}
	\avg{
		T_\chi \sigma_{ij\to ab} v}_{(T_i,T_j)}\equiv 
		& \frac {g_i} {n_{i, \mathrm{eq}} (T_i)}
	      \frac {g_j} {n_{j, \mathrm{eq}} (T_j)} 
          \int d\Pi_{i}d\Pi_{j}d\Pi_{a}d\Pi_{b}
          \times \frac {\mathbf{p}_\chi^2} {3E_\chi}\times (2 \pi)^{4}
	\nonumber\\
	&\times 
	 \delta^{(4)} (p_i + \tilde{p}_j - k_a - \tilde{k}_b)
	\times	|\mathcal{M}_{ij \to ab}|^2
	\times f_{\mathrm{eq}}(E_i,T_i)\times f_{\mathrm{eq}}(E_j,T_j).
\end{align}

The thermal averaged momentum exchange in scattering terms include   
$\mathcal{S}_{\chi\phi}$, $\mathcal{S}_{\chi\varphi}$, $\mathcal{S}_{\phi\chi}$, and 
$\mathcal{S}_{\phi\varphi}$. 
Taking $\mathcal{S}_{\chi\phi}$ as an example, 
the scattering term for $\chi(p_\chi)\phi(p_\phi)\to\chi(\tilde{p}_\chi)\phi(\tilde{p}_\phi)$ 
in Eq.~\eqref{tempBEs} can be explicitly written as 
\begin{align}
	2g_\chi
	\int d\Pi_{\chi} \frac{\mathbf{p}_\chi^2} {3E_\chi} C_{\chi\phi\to\chi\phi}
	=  {}&
	\int d\Pi_{\chi} \frac{\mathbf{p}_\chi^2} {3 E_\chi} d\Pi_{\phi} 
	d\Tilde{\Pi}_{\chi} d\Tilde{\Pi}_{\phi} 
	(2\pi)^{4} \delta^{(4)} (p_\chi + p_\phi - \tilde{p}_\chi - \tilde{p}_\phi)
	\nonumber\\
	& \times |\mathcal{M}_{\chi \phi \to \chi \phi}|^2\left[
		f_\chi(\tilde{E}_\chi,T_{\chi}) f_\phi(\tilde{E}_\phi,T_{\phi})
		- f_\chi(E_\chi,T_\chi) f_\phi(E_\phi,T_\phi)
	\right]
	\nonumber\\
	= {}& \mathcal{S}_{\chi\phi}(T_\chi,T_\phi)
	n_\chi n_\phi,
\label{eq:xphi_scatter}
\end{align}
where we have defined
\begin{align}
	\mathcal{S}_{\chi\phi}(T_\chi,T_\phi)
	\equiv &
	\frac{1}{n_{\chi, \text{eq}} (T_\chi)}
	\frac{1}{n_{\phi, \text{eq}} (T_\phi)}
	\int d\Pi_\chi d\Pi_\phi
	f_{\chi, \text{eq}} (T_\chi, E_\chi)
	f_{\phi, \text{eq}} (T_\phi, E_\phi)
	\nonumber\\
	& \times
	\int d\Tilde{\Pi}_\chi d\Tilde{\Pi}_\phi
	\left(
		\frac{\Tilde{\mathbf{p}}_\chi^2} {3\Tilde{E}_\chi}
		- \frac{\mathbf{p}_\chi^2} {3E_\chi}
	\right)
	(2\pi)^4 \delta^{(4)}
	(p_\chi + p_\phi - \Tilde{p}_\chi - \Tilde{p}_\phi)
	| \mathcal{M}_{\chi \phi \to \chi \phi} |^2
	\\
	= {} &
	\frac{g_\chi}{n_{\chi, \text{eq}} (T_\chi)}
	\frac{g_\phi}{n_{\phi, \text{eq}} (T_\phi)}
	\int_{m_\chi}^\infty dE_\chi
	\int_{m_\phi}^\infty dE_\phi
	\int_{s_\text{min}}
    	^{s_\text{max}} ds
	f_{\chi, \text{eq}} (T_\chi, E_\chi)
	f_{\phi, \text{eq}} (T_\phi, E_\phi)
	\nonumber\\
	& \times
	\frac{ \sqrt{\lambda(s, m_\chi^2, m_\phi^2)} }
	{2\times (2\pi)^4}
	\left[
    	- \frac{\mathbf{p}_\chi^2} {3E_\chi}
		\sigma_{\chi \phi \to \chi \phi} 
		+
		\int d\Omega
		\frac{d\sigma_{\chi \phi \to \chi \phi} } {d\Omega}
		\frac{\Tilde{\mathbf{p}}_\chi^2} {3\Tilde{E}_\chi}
	\right].
\label{eq:Schiphi}
\end{align}
Unlike the conventional cross-section computation performed in the center of mass frame, 
the term $\Tilde{\mathbf{p}}_\chi^2 / \Tilde{E}_\chi$ has to be given in 
the laboratory frame which discards the usual center-of-mass simplifications. 
\footnote{Comparing our results from $\mathcal{S}$ terms with those from the Fokker-Planck approach~\cite{Binder:2021bmg} in the highly non-relativistic case, we find them in good agreement.}
Comprehensive and detailed calculations are available in Appendix~\ref{sec:0th_and_2nd}. 
Due to the CPU-intensive computation required for the $\mathcal{S}$ terms, we numerically tabulate their values as a function of temperatures and present their visualization map in Appendix~\ref{sec:scattering}.

We use the set of four coupled equations \eqref{numberBEs}--\eqref{tempBEs2} to compute the main result of this work.
In Sec.~\ref{sec:result}, we will present the numerical evolution and the implications of three scenarios.

\section{Numerical result}
\label{sec:result}

In Big Bang Nucleosynthesis theory, we commonly assume that the distribution of relativistic particle number density follows a thermal equilibrium distribution. Therefore, we apply the same assumption to thermal dark matter and use it as the initial condition for the Boltzmann equation. According to the convention in Ref.~\cite{Kamionkowski:1990ni}, the non-relativistic condition is $x\equiv m_\chi/T \gg 3$, thus we evolve the Boltzmann equations with the initial conditions $T_\chi=T_\phi=T$ starting from $x=3$.

To easily demonstrate our result, we set $\mu_3=0$ to exclude the three-scalar-vertex in this study, allowing all interaction cross sections between $\chi$ and the SM to be scaled by $\sin\theta$ and $c_s$. 
In this study, we ignore $\lambda_\Phi$ to focus on the interactions among three sectors, and set $c_p=0$ for simplicity.
In the following subsections, we will present the evolution of the comoving number density $Y_i$ and temperature $y_i$ with respect to $x$ for three scenarios. For comparison, we will calculate the DM relic density using the evolution of DM density alone (denoted as ``\texttt{Only-$Y_\chi$ BE}''), DM density and temperature (denoted as ``\texttt{$Y_\chi$-and-$y_\chi$ BEs}''), and the full Boltzmann equations (denoted as ``\texttt{Full BEs}''). The method \texttt{Only-$Y_\chi$ BE} is similar to how \texttt{MicrOMEGAs} works, while \texttt{$Y_\chi$-and-$y_\chi$ BEs} would be similar to the \texttt{DRAKE} code~\cite{Binder:2021bmg}. 
Results labeled ``\texttt{Full BEs}'' correspond to our set of four coupled Boltzmann equations that are studied in this work for the first time.

\subsection{Scenario (i): Forbidden DM $m_{\phi}\thickapprox m_{\chi}$}


\begin{figure}
    \includegraphics[width=\textwidth]{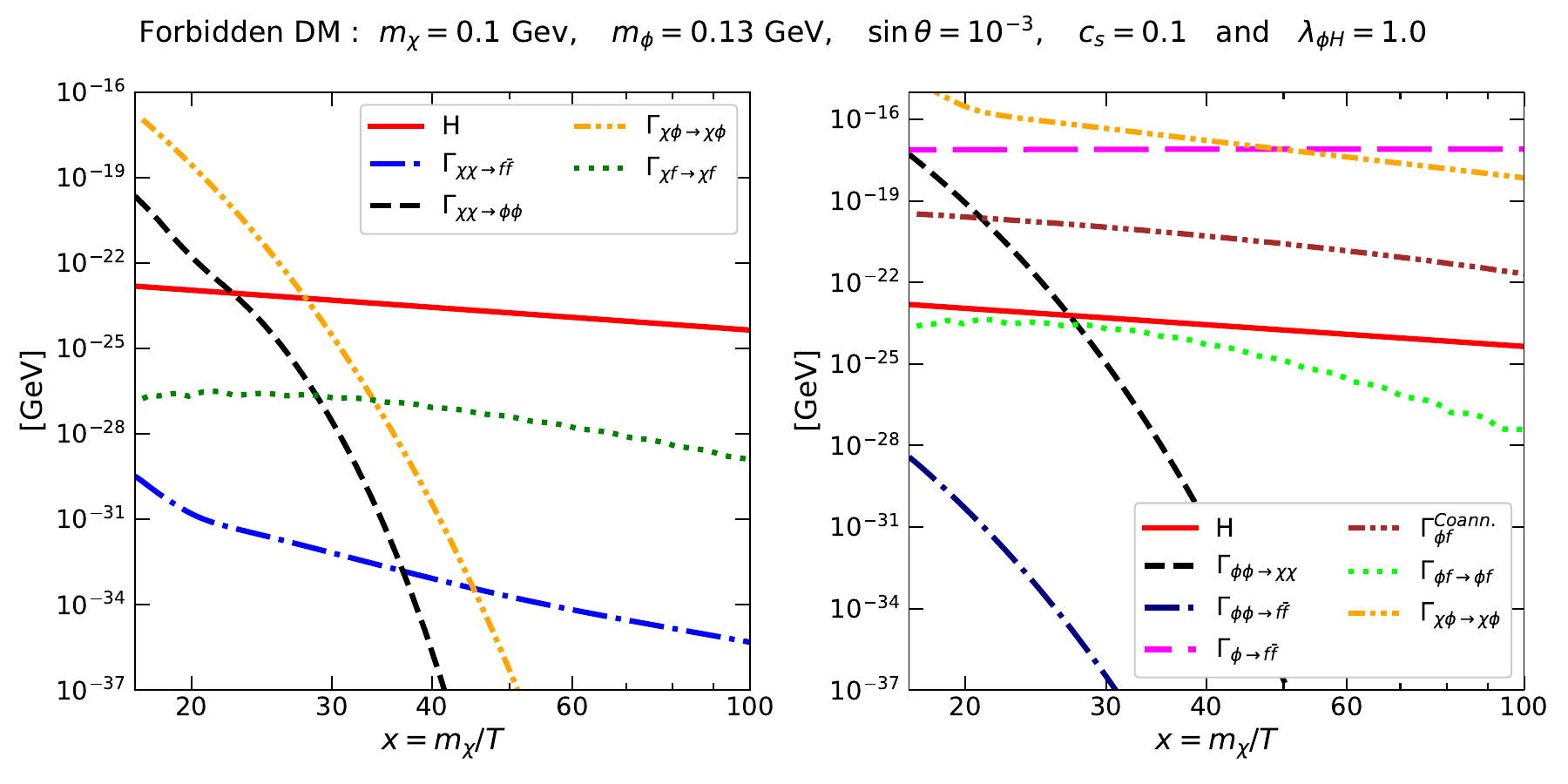}
    \caption{%
    Reaction rates for the forbidden DM scenario using $m_\chi = 0.1$~GeV, 
    $m_\phi = 0.13$~GeV, $\sin\theta = 10^{-3}$, $c_s = 0.1$ and $\lambda_{\phi H}=1.0$. 
    The left panel shows the interactions relevant to $\chi$, while 
    the interactions in the right panel correspond to $\phi$.  
    Here, we denote $f^\pm$, $g$, and $\gamma$ as charged fermions, gluon, and photon in the SM, respectively.
}
    \label{fig:forbidden_rac}
\end{figure}

In Fig.~\ref{fig:forbidden_rac}, we present the comparison of interaction rates with the Hubble parameter $H$ in the context of the forbidden DM scenario. 
Adopting $m_{\chi} = 0.1\gev$, $m_{\phi}= 0.13\gev$, $c_{s} = 0.1$ and $\lambda_{\phi H} = 1.0$ as benchmark values, we set $\sin\theta = 10^{-3}$ to catch the unique characteristics of the forbidden DM nature. In the left panel, elastic scattering $\chi \phi\to \chi \phi$ dominates the evolution at $x\lsim 30$, while annihilation $\chi\chi\to\phi\phi$ is subdominant and decouples around $x\approx 22$. This implies that kinetic decoupling between $\chi$ and $\phi$ sectors occurs after their chemical decoupling. In the right panel, the $\phi$ sector, despite $\phi\to\chi\chi$ being forbidden, maintains thermal equilibrium with the SM sector efficiently through processes like $\phi\to f\bar{f}$ and co-annihilation.

\begin{figure}[t]
    \centering
    \includegraphics[width=\textwidth]{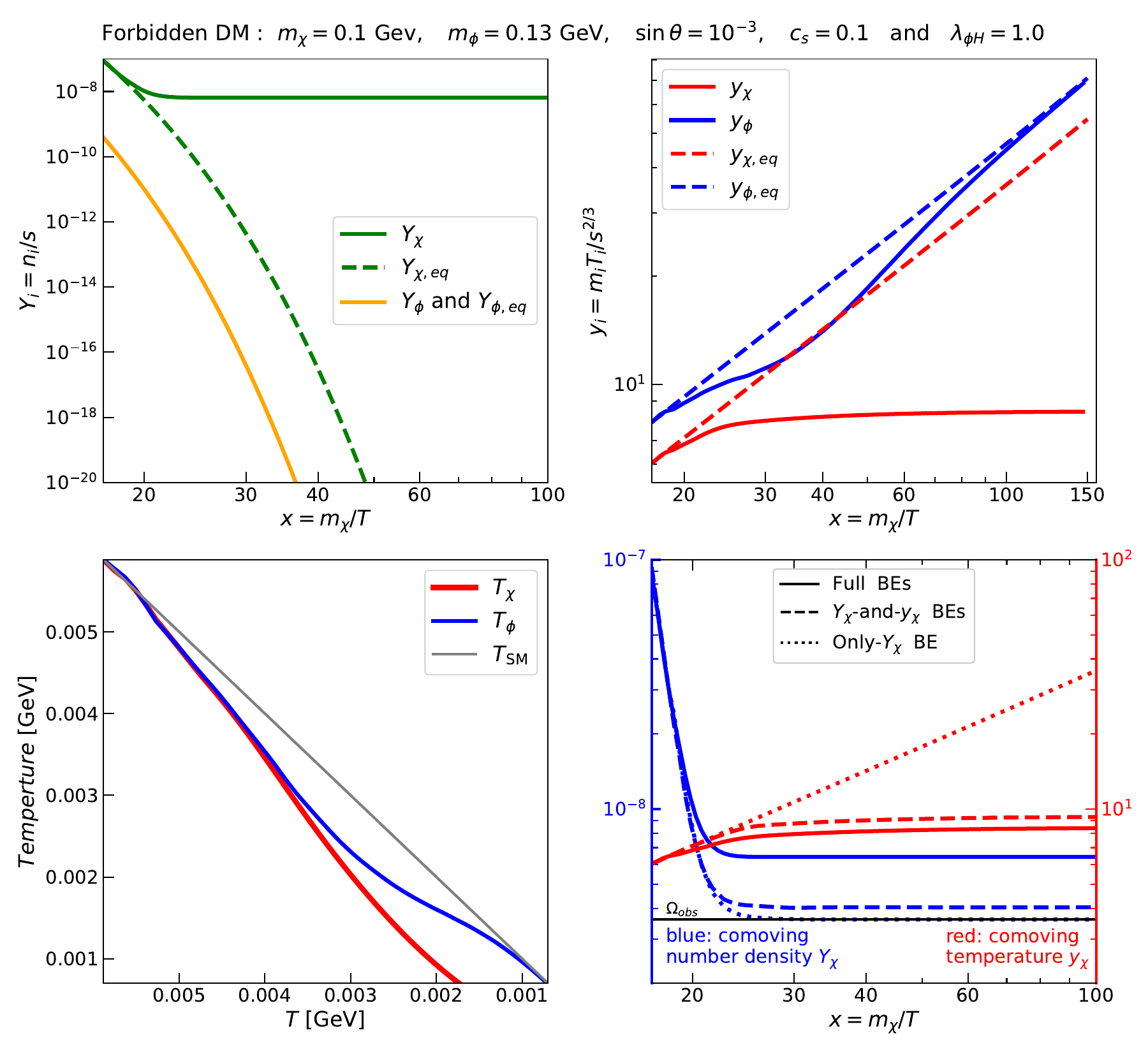}
    \caption{%
        Numerical evolution for the forbidden DM scenario by using parameters
        $m_\chi = 0.1\gev$, $m_\phi = 0.13\gev$, $\sin\theta = 10^{-3}$, $c_s = 0.1$ and $\lambda_{\phi H}=1.0$. 
        The upper left and upper right panels show the number density $Y$ and temperature $y$ evolution with solid and dashed lines representing actual numerical evolution and predicted results in thermal equilibrium. 
        In the lower-left panel, $T_\chi$ is presented with a red solid line, $T_\phi$ with a blue dashed line, and $T_{\rm SM}$ with a gray dash-dotted line while varying $T_{\rm SM}$. The lower right panel illustrates comoving number density $Y_\chi$ (blue left axis) and comoving temperature $y_\chi$ (red right axis) using blue and red lines; 
        Those solid, dashed, and dotted lines are based on the calculation of full Boltzmann equations, $Y_\chi$ and $y_\chi$ Boltzmann equations, and only $Y_\chi$ Boltzmann equation, respectively.
        }
    \label{fig:forbidden_ev}
\end{figure}

In Fig.~\ref{fig:forbidden_ev}, we illustrate the evolution obtained by numerically solving Eq.~\eqref{eq:Yy}
for the comoving number density $Y_{i}$ (upper left), comoving temperature $y_{i}$ (upper right), $T_{\chi,\phi,{\rm SM}}$ (lower left), and a comparison of three cases
(\texttt{Only-$Y_\chi$ BE}, \texttt{$Y_\chi$-and-$y_\chi$ BEs} and \texttt{Full BEs} cases discussed before)
in the lower right panel.

The upper-left panel of Fig.~\ref{fig:forbidden_ev} displays the evolution of the comoving number densities of $\chi$ and $\phi$, represented by green and orange solid lines, respectively. The dashed line corresponds to the thermal equilibrium distribution. When $Y_\chi$ (green solid line) departs from $Y_{\chi,eq}$ (green dashed line), the chemical decoupling of $\chi$ occurs at $x \approx 20$. For $x\lsim 20$, $Y_\phi$ and $Y_{\phi,eq}$ evolve similarly, due to substantial $\phi$ decay that sustains the thermal equilibrium, as illustrated in Fig.~\ref{fig:forbidden_rac}. The upper-right panel of Fig.~\ref{fig:forbidden_ev} interestingly shows the $\phi$ comoving temperature evolution. Here, $y_\phi$ (blue solid line) deviates from $y_{\phi,eq}$ (blue dashed line) at the DM freeze-out and closely, or even identically, aligns with $y_{\chi,eq}$ (red dashed line) in the $20<x<50$ range, a result of the efficient $\phi\phi\to\chi\chi$ annihilation and elastic scattering $\chi \phi\to \chi \phi$. 
However, once the $\phi\phi\to\chi\chi$ annihilation and $\chi \phi\to \chi \phi$ cease, co-annihilation take over, transferring energy between the $\phi$ and SM sectors. Consequently, $y_\phi$ becomes to $y_{\phi,eq}$ for $x>100$. The complex evolution of the $\phi$ temperature can be further understood by comparing $T_\chi$ (red line), $T_\phi$ (blue line), and $T_{\rm SM}$ (gray line) in the lower-left panel of Fig.~\ref{fig:forbidden_ev}.

From the upper-left and upper-right panels of Fig.~\ref{fig:forbidden_ev}, we observe that $y_{\phi}$ deviates from $y_{\phi,eq}$, while $Y_{\phi}$ remains close to $Y_{\phi,eq}$. This is due to the non-zero chemical potential $\mu$ required in the forbidden DM scenario. Although thermal equilibrium between $\phi$ and $\varphi$ is disrupted, their chemical equilibrium is maintained. This is further illustrated in the left panel of Fig.~\ref{fig:forbidden_rac}. For heat exchange, the dominant reactions are elastic collisions between $\phi$ and $\chi$, followed by interactions with $\varphi$. In contrast, number density exchanges mainly occur between $\phi$ and $\varphi$. Consequently, $\phi$ achieves thermal equilibrium with $\chi$ and chemical equilibrium with $\varphi$. A common thermal equilibrium among all three sectors is established when the heat exchanges between $\phi$ and $\chi$, and between $\phi$ and $\varphi$, are comparable. As the elastic scattering rate between $\chi$ and $\phi$ decreases below that of the $\phi$-$\varphi$ interactions, $\phi$ decouples from the $\chi$ sector and gradually re-establishes equilibrium with $\varphi$, with this transition occurring around $x \sim 50$.

The lower-right panel of Fig.~\ref{fig:forbidden_ev} compares numerical evolution equations for $Y$ (blue, left axis) and $y$ (red, right axis) using three computational methods: \texttt{Full BEs} (solid lines), \texttt{$Y_\chi$-and-$y_\chi$ BEs} (dashed lines), and \texttt{Only-$Y_\chi$ BE} (dotted lines). 
We first tune the coupling parameters to match the Planck measured relic density (black line) using the \texttt{Only-$Y_\chi$ BE} approach, then apply these parameters to compute the evolution with the other two methods.
Surprisingly, in the forbidden DM scenario, the \texttt{$Y_\chi$-and-$y_\chi$ BEs} approach shows little difference compared to the conventional \texttt{Only-$Y_\chi$ BE} approach, while the \texttt{Full BEs} approach yields significantly different results compared with the other two simplified approaches.

\begin{figure}[t]
    \centering
    \includegraphics[width=\textwidth]{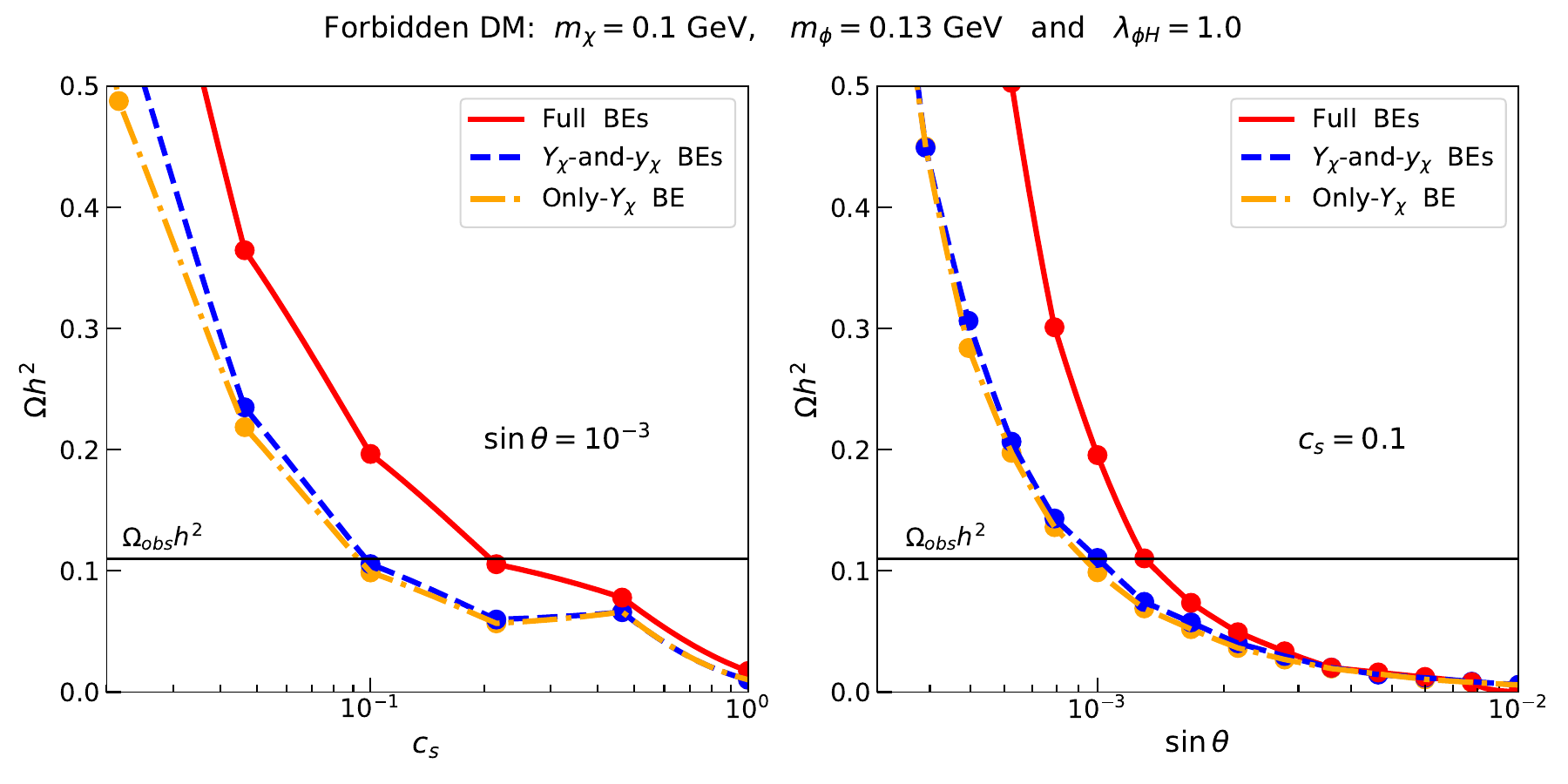}
    \caption{%
    Relic density  $\Omega_{\chi} h^2$ for the forbidden DM scenario, as a function of $c_{s}$ (left panel) and $\sin\theta$ (right panel). 
    Three numerical approaches are compared by red solid lines (\texttt{Full BEs}), 
    blue dashed lines (\texttt{$Y_{\chi}$-and-$y_{\chi}$ BEs}), and 
    orange dash-dotted lines (\texttt{Only-$Y_{\chi}$ BEs}). 
    }
    \label{fig:forbidden_relic}
\end{figure}

In Fig.~\ref{fig:forbidden_relic}, we present the relic density $\Omega_{\chi} h^{2}$ with respect to $c_s$ (left panel) and $\sin{\theta}$ (right panel). The benchmark parameters are $m_{\chi} = 0.1 \gev$, $m_{\phi} = 0.13 \gev$ and $\lambda_{\phi H}=1.0$. 
The red solid lines, blue dashed lines, and orange dash-dotted lines correspond to the \texttt{Full BEs}, \texttt{$Y_{\chi}$-and-$y_{\chi}$ BEs}, 
and \texttt{Only-$Y_{\chi}$ BEs} approaches, respectively. 
In the left panel, $\sin{\theta}=10^{-3}$ remains constant as $c_s$ changes, while in the right panel, $\sin{\theta}$ varies for $c_s=0.1$.
Both panels exhibit similar trends when $c_s$ and $\sin{\theta}$ are varied. 
The results from \texttt{$Y_{\chi}$-and-$y_{\chi}$ BEs} and \texttt{Only-$Y_{\chi}$ BEs} are similar, but those from \texttt{Full BEs} show a greater deviation due to early kinetic decoupling effects in the $\phi$ sector, which are not considered in the former methods.
Specifically, for this benchmark point ($\sin{\theta}=10^{-3}$, $c_s=0.1$), the relic density computed by \texttt{Only-$Y_{\chi}$ BEs} matches the Planck measurement $\Omega_{\chi} h^{2}=0.11$, whereas the results from \texttt{$Y_{\chi}$-and-$y_{\chi}$ BEs} and \texttt{Full BEs} exceed this value by approximately $10\%$ and $72\%$, respectively. 
Furthermore, when comparing the \texttt{Full BEs} approach with the other two approaches in both panels, 
we observe larger differences between \texttt{Full BEs} and \texttt{\texttt{Only-$Y_{\chi}$ BEs}} when varying $\sin{\theta}$ compared to varying $c_s$. 
This occurs because the interactions between $\phi$ and SM particles are proportional to $\sin^2\theta$, and affect the early kinetic decoupling of the $\phi$ sector.

\subsection{Scenario (ii): Resonance DM $m_{\phi}\thickapprox 2 m_{\chi}$}

\begin{figure}[ht]
    \centering
    \includegraphics[width=\textwidth]{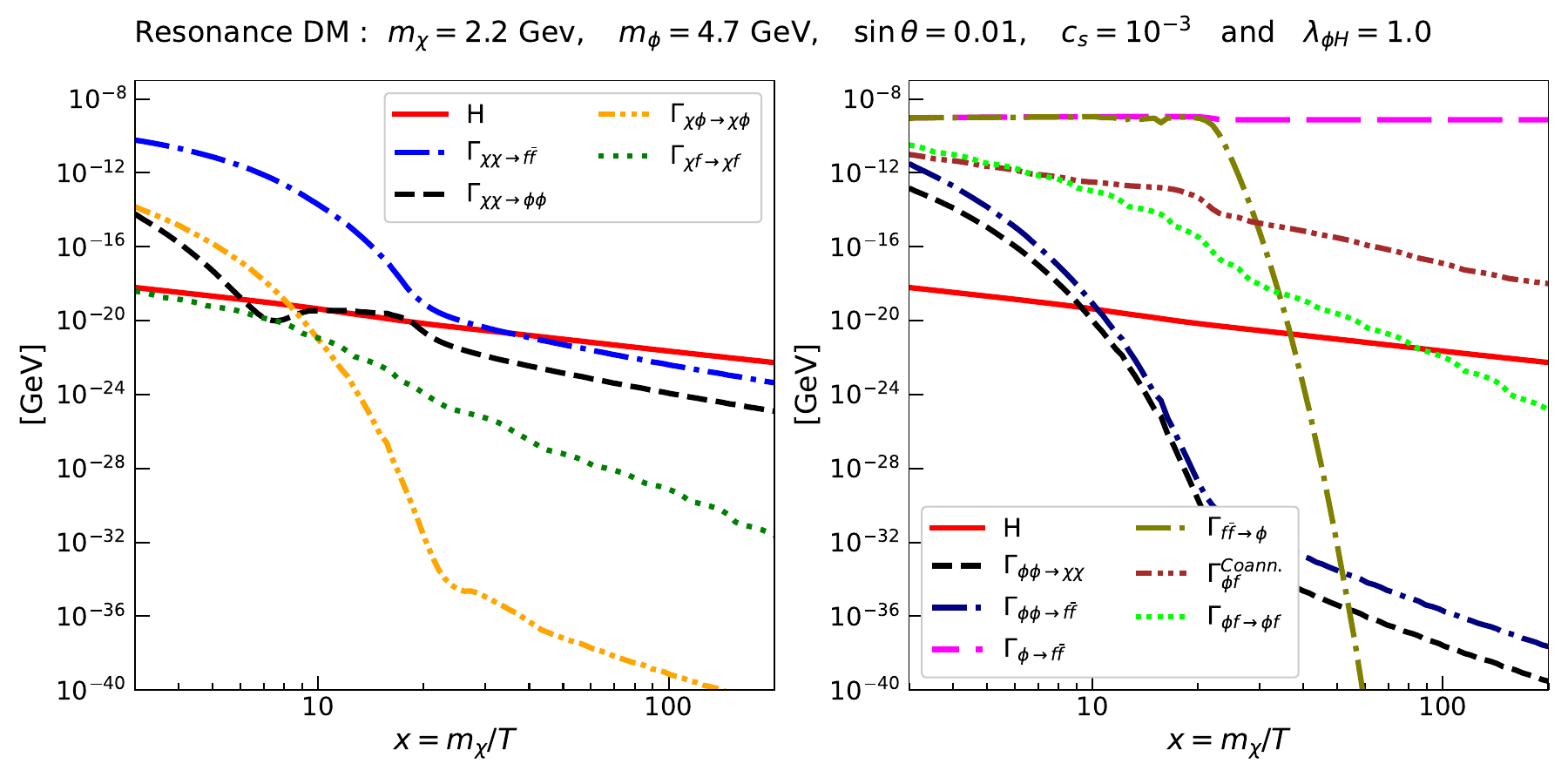}
    \caption{%
    Reaction rates for the resonance DM scenario using  $m_\chi = 2.2\gev$,
    $m_\phi = 4.7\gev$, $\sin{\theta} = 0.01$, $c_s = 10^{-3}$ and $\lambda_{\phi H}=1.0$. 
    The color scheme matches that of Fig.~\ref{fig:forbidden_rac}.}
    \label{fig:resonance_rac}
\end{figure}

For convenience, in this scenario we use $x_{\rm pole} \sim 1.5/ \left ( 1-R_\chi^2\right)$, where $R_\chi=2m_\chi/m_\phi$, to characterize the cross section peak for annihilation $\chi \chi \to f \bar{f}$. 
To emphasize the impact of temperature, we require the peak to fall within $3< x \le 25$, and choose benchmark parameters: $m_\chi = 2.2\gev$, $m_\phi = 4.7\gev$, $\sin\theta = 10^{-2}$, $c_s = 10^{-3}$ and $\lambda_{\phi H} = 1.0$.
With this setup, the characteristic peak occurs at $x\approx10$.

\begin{figure}
    \centering
    \includegraphics[width=\textwidth]{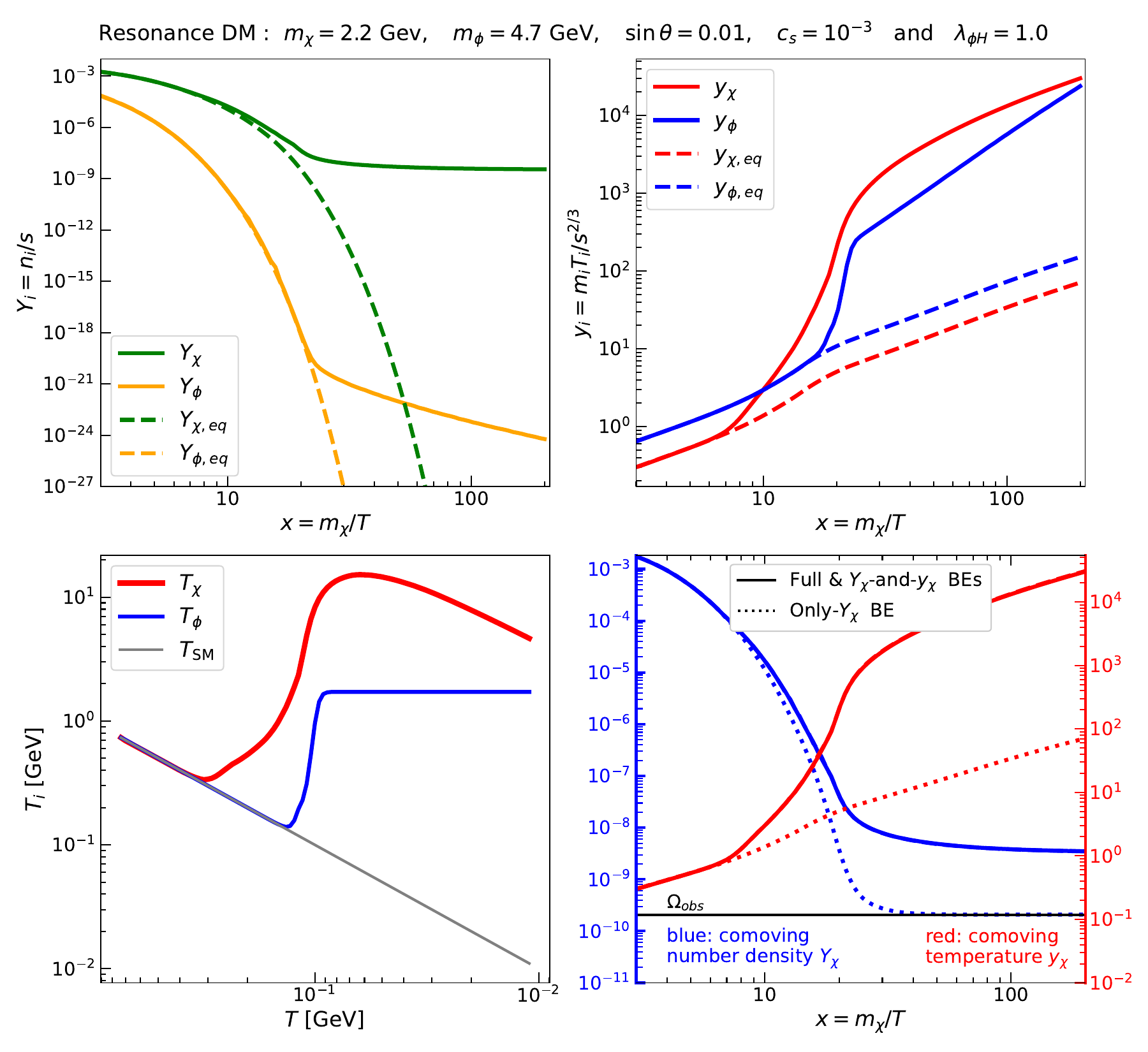}
    \caption{%
        Numerical results for the resonance DM scenario using  $m_\chi = 2.2\gev$,
        $m_\phi = 4.7\gev$, $\sin\theta = 0.01$, $c_s = 10^{-3}$ and $\lambda_{\phi H} = 1.0$.
        The color schemes are the same as Fig.~\ref{fig:forbidden_ev}.
        }
    \label{fig:resonance_ev}
\end{figure}

In Fig.~\ref{fig:resonance_rac}, we compare interaction rates with the Hubble parameter $H$ in the resonance DM scenario. Annihilation $\chi \chi \to f \bar{f}$ dominates evolution at $x\lesssim 20$ in the left panel, while the elastic scattering rate $\chi f \to \chi f$ stays below $H$, indicating that heat transfer between $\chi$ and SM particles relies solely on annihilation. Due to the mass condition $m_\phi\approx 2m_\chi$, $\chi\chi \to \phi \phi$ annihilation and $\chi\phi$ elastic scattering decouple at $x\approx 6$ and $x\approx 8$, respectively. 
This suggests kinetic decoupling between $\chi$ and $\phi$ sectors, occurring before the DM freeze-out. The reaction rate $\chi \chi \to \phi \phi$ exhibits a dip at $x=10$ due to rapid reduction of $\chi$ near the peak of $\chi \chi \to f \bar{f}$, something that will be discussed further when we introduce Fig.~\ref{fig:resonance_ev}. In the right panel, $\phi$ efficiently maintains thermal equilibrium with the SM via $\phi \to f \bar{f}$. Additionally, at $x\approx 23$, $\phi \to f \bar{f}$ significantly exceeds $f \bar{f} \to \phi$, representing heat transfer solely from $\phi$ to the SM sector.

We illustrate the resulting numerical evolution of the resonance scenario in Fig.~\ref{fig:resonance_ev}. In the upper-left panel of Fig.~\ref{fig:resonance_ev}, when $Y_{\chi}$ (green solid line) and $Y_{\phi}$ (orange solid line) depart from $Y_{\chi,\text{eq}}$ (green dashed line) and $Y_{\phi,\text{eq}}$ (orange dashed line), the freeze-out of $\chi$ and $\phi$ occurs at $x\approx20$ and $x\approx23$, respectively. 
The upper-right panel shows that $y_{\chi}$ (red solid line) departs from $y_{\chi,\text{eq}}$ (red dashed line) at $x\approx7$, and $y_{\phi}$ (blue solid line) leaves $y_{\phi,\text{eq}}$ (blue dashed line) at $x\approx18$. 
This indicates that for both $\chi$ and $\phi$, their kinetic decoupling appears before their chemical decoupling.

From the low-left panel in Fig.~\ref{fig:resonance_ev}, we can observe that the three sectors maintain thermal equilibrium initially, then $T_{\chi}$ and $T_{\phi}$ increase rapidly until $T_{\rm SM}\approx 0.1\gev$. However, in the range $T_{\rm SM}\gsim 0.1\gev$, $T_{\phi}$ is flat but $T_\chi$ decreases. 
From looking at Fig.~\ref{fig:resonance_rac}, we can see that the increase in $T_{\chi}$ results from $\chi \chi \to f \bar{f}$ reaching the cross-section peak and, therefore,
DM particles with temperature $T_\chi$ being massively consumed while new DM particles with higher speed are continuously generated by $\phi$-decay.
To all that, adding that the higher temperature enhances the averaged cross-section, thus, the intriguing $\Gamma_{\chi \chi \to \phi \phi}$ dip in Fig.~\ref{fig:resonance_rac} can be related to the decrease in $n_{\chi}$ first accompanied by  
the successive enhancement of averaged cross-section. 
Similarly, $T_{\phi}$ increases because $f \bar{f} \to \phi$ 
becomes less efficient than $\phi \to f \bar{f}$ with $\phi$s with lower temperature being massively consumed while new $\phi$s with higher speed are generated by $\phi f\to \phi f$ and those inverse co-annihilation processes.

In the lower-right panel of Fig.~\ref{fig:resonance_ev}, the \texttt{Full BEs} and \texttt{$Y_{\chi}$-and-$y_{\chi}$ BEs} approaches yield almost identical results, contrasting significantly with those from using \texttt{Only-$Y_{\chi}$ BEs}. 
The \texttt{Full BEs} approach agrees with the one obtained using \texttt{$Y_{\chi}$-and-$y_{\chi}$ BEs} because $\phi$ sector remains in equilibrium with the SM sector even after $\chi$-SM decoupling. However, the difference with \texttt{Only-$Y_{\chi}$ BEs} approach arises as $\chi$ kinetically decouples from the other two sectors before its chemical decoupling. 
Hence, for this benchmark resonance scenario, the standard calculation using \texttt{Only-$Y_{\chi}$ BEs} may underestimate the relic density by a factor of ten.

\begin{figure}[t]
    \centering
    \includegraphics[width=\textwidth]{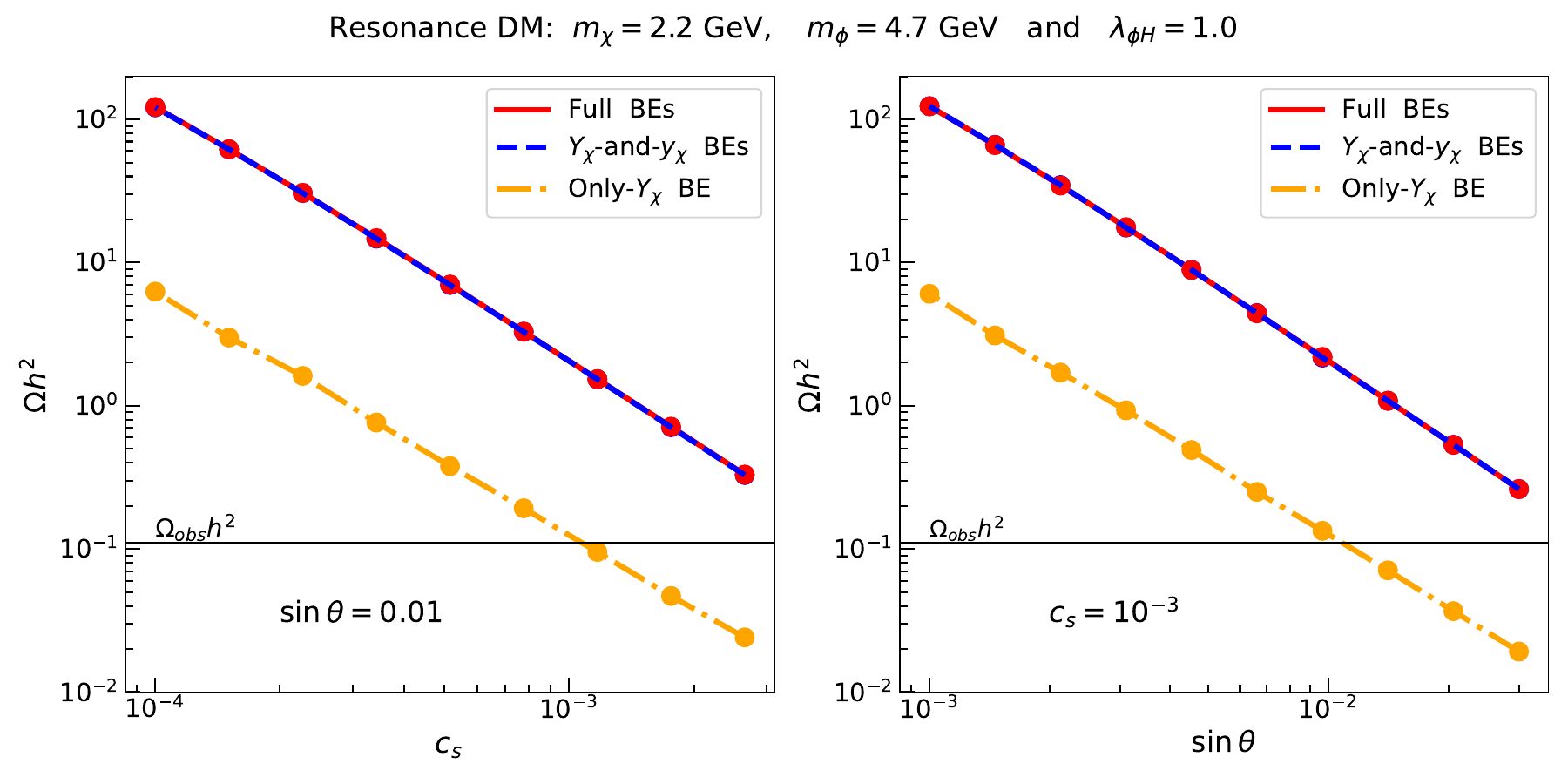}

    \caption{
    Same as Fig.~\ref{fig:forbidden_relic} but for the resonance DM scenario with parameters $m_\chi=2.2\gev$ and $m_\phi=4.7\gev$. 
    }
    \label{fig:resonance_relic}
\end{figure}

In Fig.~\ref{fig:resonance_relic}, we depict the relic density $\Omega_{\chi} h^{2}$ as a function of $c_s$ (left panel) and $\sin{\theta}$ (right panel), with $m_{\chi} = 2.2 \gev$ and $m_{\phi} = 4.7 \gev$ as benchmark points. Similar to Fig.~\ref{fig:forbidden_relic}, the left panel uses fixed $\sin{\theta}=0.01$ with different $c_s$ values, while the right panel has fixed $c_s=10^{-3}$ with changing $\sin{\theta}$ to calculate the relic density.
\textit{Overall, both panels show that results from \texttt{Full BEs} and \texttt{$Y_{\chi}$-and-$y_{\chi}$ BEs} are consistent but differ by approximately an order of magnitude from the result using \texttt{Only-$Y_{\chi}$ BEs}, for all the values of $c_s$ and $\sin\theta$ displayed in Fig.~\ref{fig:resonance_relic}}.
Referring back to Fig.~\ref{fig:resonance_rac}, we note that the heat transfer between the dark and SM sectors depends entirely on the process $\chi \chi \to f \bar{f}$, with heat and density transfer being controlled by $\sin\theta$ and $c_s$ in the same way. 
Therefore, altering $\sin\theta$ and $c_s$ cannot induce kinetic decoupling of DM following its chemical decoupling, 
as heat transfer is always less efficient than number density transfer in the $\chi \chi \to f \bar{f}$ process.
After kinetic decoupling, the heat transfer between the dark sector and the other two sectors is trivial, and the relic density is inversely proportional to the annihilation cross-section $\langle\sigma v(\chi \chi \to f \bar{f})\rangle\propto (c_s \sin\theta)^2$. 
Thus, in contrast with the forbidden scenario, the three numerical approaches in the resonance scenario cannot be unified at large $c_s$ and $\sin\theta$ values.

We would like to address the case where the $\phi$ sector decouples from the SM sector before the dark sector, potentially leading to different outcomes between \texttt{Full BEs} and \texttt{$Y_{\chi}$-and-$y_{\chi}$ BEs}. 
However, finding a parameter combination where such case is realized is challenging. 
A weaker interaction causing earlier decoupling of the $\phi$ sector demands a smaller $\sin\theta$ and a finely tuned value for the resonance parameter ($1-R_\chi^2$). 
The annihilation process $\chi \chi \to f \bar{f}$ with these parameters may reach its peak later than our intended time frame, leading to an overabundance of relic density. 
Moreover, maintaining thermal equilibrium among the three sectors at $x\approx3$, our default initial conditions, becomes difficult with such a small $\sin\theta$. Therefore, we can conclude that in the resonance scenario, it is crucial to precisely consider the evolution of DM temperature, without extending to include $\phi$ evolution.

\subsection{Scenario (iii): Secluded DM: $m_{\phi}\ll m_{\chi}$}

\begin{figure}[ht]
    \centering
    \includegraphics[width=\textwidth]{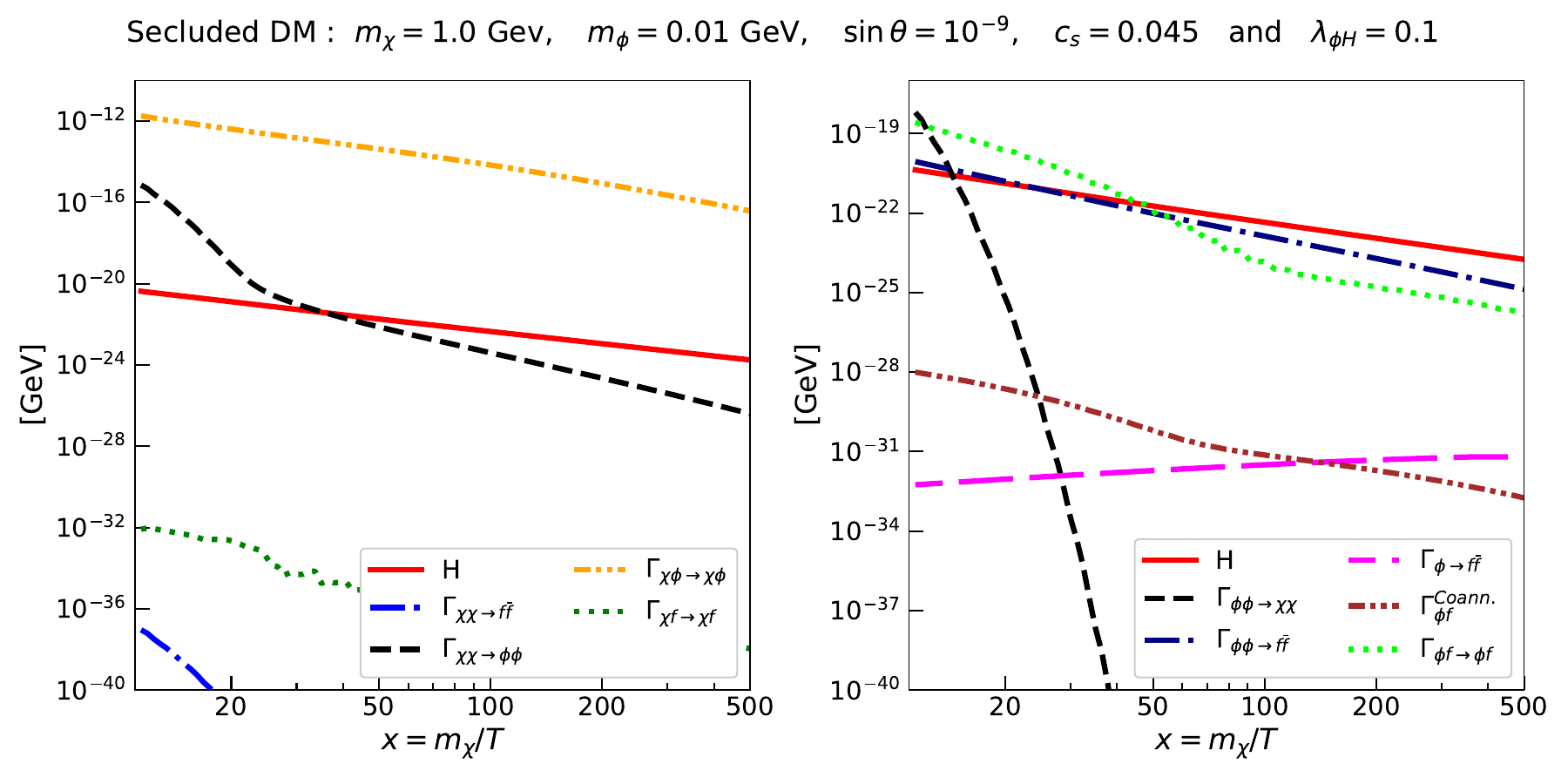}
    \caption{%
    Comparison of interaction rates with the Hubble parameter $H$ in the context of the secluded DM scenario,  
    by using model parameters are $m_\chi = 1.0$\gev, $m_\phi = 0.01$\gev,  $\sin\theta = 10^{-9}$, $c_s = 0.045$ and $\lambda_{\phi H} = 0.1$.
    The color scheme follows Fig.~\ref{fig:forbidden_rac}.}
    \label{fig:secluded_rac}
\end{figure}

Lastly, we examine the secluded DM scenario in detail. 
In this scenario, the main annihilation channel of the secluded DM scenario is $\chi\chi\to\phi\phi$.
In our setup, by taking $c_p=0$, annihilation $\chi\chi\to\phi\phi$ exclusively through $c_s$ results in purely $p$-wave contribution.
To obtain the correct relic density, the cross-section for this annihilation must have a large phase space integral, particularly when $m_\chi \gg m_\phi$, to compensate for tiny $\sin{\theta}$. 
Hence, we set $\sin{\theta}=10^{-9}$, $m_{\chi} = 1.0\gev$, $m_{\phi}= 0.01\gev$, $c_{s} = 0.045$, and $\lambda_{\phi H} = 0.1$ as benchmark parameters for our analysis of the secluded DM scenario.

In Fig.~\ref{fig:secluded_rac}, we present the comparison of interaction rates against $H$ in the context of the secluded DM scenario. 
The left panel shows scatterings related to $\chi$,
where we see that the evolution of $\chi$ is primarily driven by the highly efficient elastic scattering $\chi \phi \to \chi \phi$, while the subdominant process $\chi \chi \to \phi \phi$ decouples at $x\approx22$. 
This indicates that kinetic decoupling between $\chi$ and $\phi$ occurs after their chemical decoupling. 
Regarding interactions between $\chi$ and the SM sector, elastic scatterings $\chi f \to \chi f$ and annihilations $\chi \chi \to f \bar{f}$ are far below the Hubble parameter due to suppression by $\sin{\theta}$, suggesting no direct heat transfer between $\chi$ and the SM sector, but rather indirect transfer via $\phi$. 
In the right panel, it can be observed that the dominant process is $\phi f\to \phi f$, decoupling at $x\approx50$, while the subdominant reaction rate for $\phi \phi \to f \bar{f}$ falls an order of magnitude below $H$ for $x>100$. 
Regarding $\phi$ decay and co-annihilation, their reaction rates remain significantly below $H$, indicating negligible effects in the evolution of $\phi$.

\begin{figure}
    \centering
    \includegraphics[width=\textwidth]{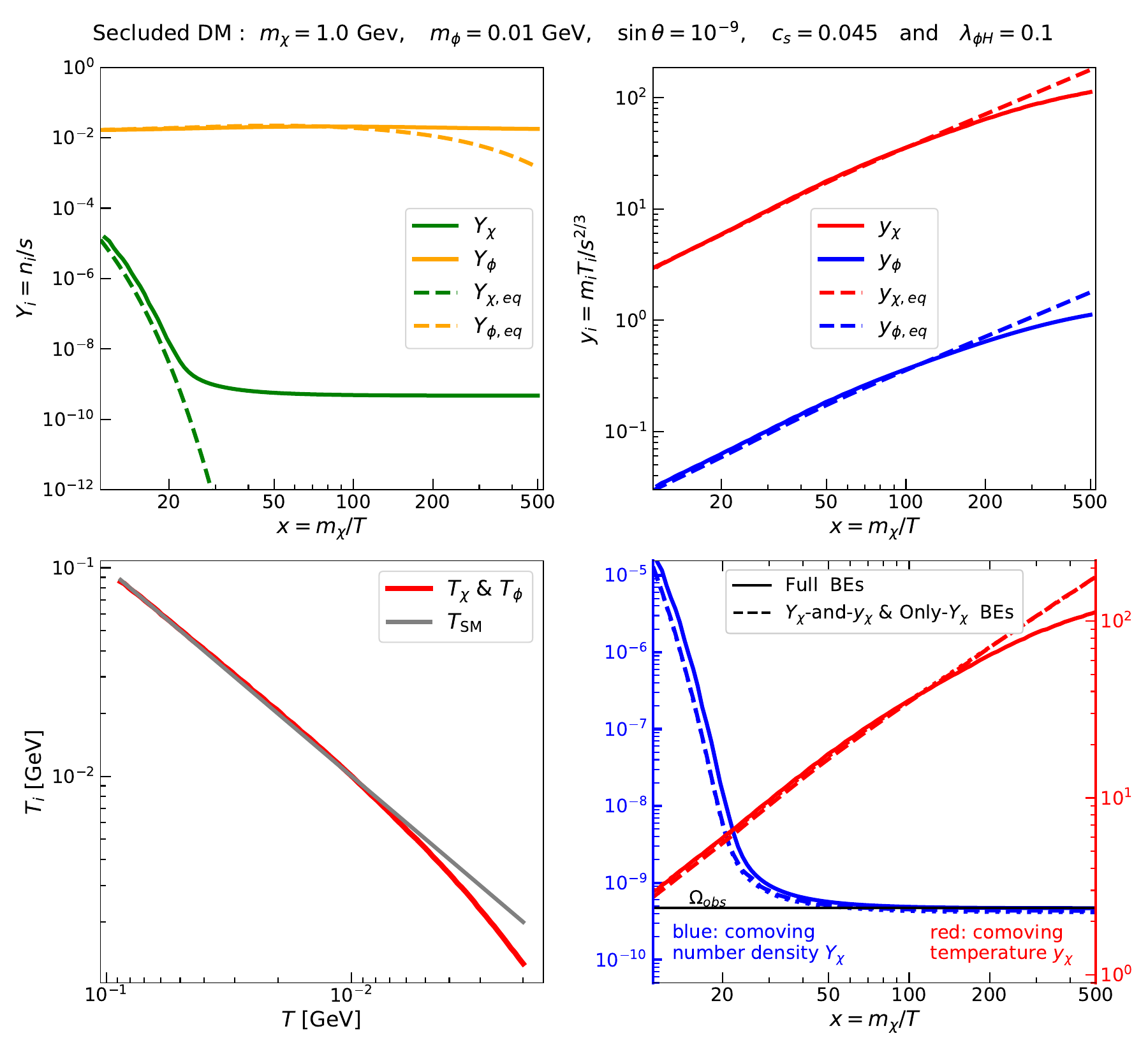}
    \caption{%
        Same as on Fig.~\ref{fig:forbidden_ev} but showing for the secluded DM with $m_\chi = 1.0$\gev, 
        $m_\phi = 0.01$\gev,  $\sin{\theta} = 10^{-9}$, $c_s = 0.045$ and $\lambda_{\phi H} = 0.1$. 
        }
    \label{fig:secluded_ev}
\end{figure}

The numerical evolution results of the secluded scenario are illustrated in Fig.~\ref{fig:secluded_ev}. 
In the upper-left panel, both $Y_{\chi}$ (green solid line) and $Y_{\phi}$ (orange solid line) deviate from their equilibrium values ($Y_{\chi,\text{eq}}$ in the green dashed line and $Y_{\phi,\text{eq}}$ in the orange dashed line), marking the freeze-out of $\chi$ and $\phi$ at approximately $x\approx 22$ and $x\approx 120$, respectively. 
In the upper-right panel, both $y_{\chi}$ (red solid line) and $y_{\phi}$ (blue solid line) differ from their equilibrium values ($y_{\chi,\text{eq}}$ in red dashed line and $y_{\phi,\text{eq}}$ in blue dashed line) around $x\approx 150$. 
This implies that both $\chi$ and $\phi$ undergo kinetic decoupling later than their chemical decoupling.

The temperature evolution among the three sectors, illustrated in the lower-left of Fig.~\ref{fig:secluded_ev}, highlights that the temperatures of $\chi$ and $\phi$ remain identical within our range of interests due to efficient heat transfer via elastic scattering $\chi \phi \to \chi \phi$, until they deviate from $T_{\rm SM}$ around $T \approx 0.006\,\text{GeV}$ when $\phi \phi \to f \bar{f}$ decouples. 
This shared temperature between the dark and $\phi$ sectors, together with the kinetic decoupling of $\chi$ occurring after its chemical decoupling, results in nearly identical outcomes between the \texttt{$Y_{\chi}$-and-$y_{\chi}$ BEs} and \texttt{Only-$Y_{\chi}$ BEs} methods, shown as dashed lines in the lower-right panel of Fig.~\ref{fig:resonance_ev}. 
They are slightly different from \texttt{Full BEs} due to asymmetry in the $\mathcal{S}_{\chi\phi}$ term\footnote{Refer to Fig.~\ref{fig:schiphi} in Appendix~\ref{sec:scattering}}. 
The asymmetry disappears when the masses of the incoming particles are of similar order (see the first two figures in Fig.~\ref{fig:schiphi} in Appendix~\ref{sec:scattering}). In Fig.~\ref{fig:secluded_rac}, the direct interaction between $\chi$ and $\varphi$ is decoupled early on, so $\chi$ only exchanges heat with $\varphi$ indirectly via $\phi$. The symmetry of $\mathcal{S}_{\phi\varphi}$ arises from the relativistic nature of $\phi$ and $\varphi$, but $\mathcal{S}_{\chi\phi}$ is asymmetric due to the non-relativistic $\chi$. These two factors together lead to deviations of $T_{\chi}$ from $T$, as well as deviations of \texttt{Full BEs} from \texttt{Only-$Y_{\chi}$ BEs}.

Nevertheless, this difference leads to a significant systematic uncertainty of around $9\%$, comparable to uncertainties originating from the entropy table~\cite{Drees:2015exa} but surpasses the uncertainty of Planck measurement~\cite{Aghanim:2018eyx}.

Similar to the DM resonance case, in this scenario we did not find a suitable range for $c_s$ and $\sin\theta$ where the \texttt{Full BEs} approach would show a significant shift. 
In our secluded DM setup, where the thermal equilibrium is maintained between the dark and $\phi$ sectors, and all three sectors are assumed to be in thermal equilibrium at $x=3$, the improvement brought by the \texttt{Full BEs} approach is not as pronounced as in the forbidden DM scenario. 
Furthermore, the discrepancy arising from the asymmetry in the $\mathcal{S_{\chi\phi}}$ term must be sizable, because secluded DM particles mainly annihilate via the $\chi\chi\to\phi\phi$ channel controlled by the same couplings as $\chi\phi\to\chi\phi$. 
Additionally, $\sin\theta$ has no impact on the $\mathcal{S_{\chi\phi}}$ term.
Therefore, it is unnecessary to include plots analogous to Figs.~\ref{fig:forbidden_relic} and \ref{fig:resonance_relic} in this scenario.

\section{Summary and conclusion}
\label{sec:conclusion}

When the dark sector interacts with the SM sector through a light scalar particle $\phi$, the heat transfer between the dark and SM sectors becomes intricate. 
Previous studies often rely on the simplest assumption to calculate the DM relic density: these two sectors are in thermal equilibrium before freeze-out. 
However, when incorporating the interaction between $\phi$ and the SM sector, the mixing angle $\sin\theta$ is strongly constrained by experimental data, leading to insufficient heat transfer between $\phi$ and SM particles to maintain thermal equilibrium. 
Moreover, the absence of a signal in DM direct detection experiments may indicate that the coupling between DM and $\phi$ ($c_s$ in the present analysis) may be suppressed. 
Consequently, three distinct sectors emerge---the DM sector, the $\phi$ sector, and the SM sector---prior to the DM freeze-out. 
It becomes important to consider Boltzmann equations for the densities and temperatures of all relevant sectors to accurately compute the DM relic density. 
In this study, we have developed four fully coupled Boltzmann equations to precisely determine the relic density and dark temperature with contributions from these three sectors.

We utilize the minimal Higgs portal Model to study Boltzmann equations for densities and temperatures, focusing on light $\chi$ and $\phi$ particles with masses approximately $\mathcal{O}(1)\gev$ or lower. The relevant model parameters include $m_\chi$, $m_\phi$, $\sin\theta$, $c_s$ and $\lambda_{\phi H}$, governing the interaction rate. Inspired by previous works, we investigate three scenarios featuring early decoupling: (i) forbidden DM with $m_\chi\lesssim m_\phi$, (ii) resonance DM with $2 m_\chi\approx m_\phi$, and (iii) secluded DM with $m_\chi\gg m_\phi$; crucial for understanding DM annihilation in the early universe. 
We analyze the evolution of comoving number density $Y_i$ and temperature $y_i$ for these scenarios. 
To quantitatively verify the results obtained by the full Boltzmann equations (\texttt{Full BEs}), we compare them against other well studied approaches: considering only DM density evolution (\texttt{Only-$Y_\chi$ BE}), and incorporating both DM density and temperature evolutions (\texttt{$Y_\chi$-and-$y_\chi$ BEs}). 
While these two approaches resemble existing codes like \texttt{MicrOMEGAs} and \texttt{DRAKE}, the approach of this work, \texttt{Full BEs}, is introduced for the first time.

In the forbidden DM scenario, while the results from \texttt{Only-$Y_{\chi}$ BEs} and \texttt{$Y_{\chi}$-and-$y_{\chi}$ BEs} show similarity, the \texttt{Full BEs} method exhibits a more significant deviation. 
This is due to early kinetic decoupling effects within the $\phi$ sector. 
Specifically, using a benchmark point we show that 
the relic density calculated by \texttt{Only-$Y_{\chi}$ BEs} could align with Planck measurements, 
while the results from \texttt{$Y_{\chi}$-and-$y_{\chi}$ BEs} and \texttt{Full BEs} could exceed this value by approximately $10\%$ and $72\%$, respectively. 
When varying $\sin{\theta}$ compared to varying $c_s$, larger differences are observed between \texttt{Full BEs} and \texttt{Only-$Y_{\chi}$ BEs} 
because of the early kinetic decoupling in the $\phi$ sector.

In the resonance DM scenario, heat transfer compared to number density transfer is less efficient. Moreover, modifying $\sin\theta$ and $c_s$ does not lead to kinetic decoupling of DM after chemical decoupling, as long as we require the resonance peak to happen at a higher temperature than the freeze-out temperature.
Unlike the forbidden scenario, the three numerical approaches in the resonance scenario cannot be reconciled at large $c_s$ and $\sin\theta$. 
Results from \texttt{Full BEs} and \texttt{$Y_{\chi}$-and-$y_{\chi}$ BEs} remain consistent but deviate by an order of magnitude from \texttt{Only-$Y_{\chi}$ BEs}, regardless of $c_s$ and $\sin\theta$ values.

In the secluded DM scenario, the outcomes of the \texttt{$Y_{\chi}$-and-$y_{\chi}$ BEs} and \texttt{Only-$Y_{\chi}$ BEs} methods are nearly identical due to the nature of the scenario. However, the \texttt{Full BEs} approach can introduce a considerable systematic uncertainty of approximately 
$9\%$, comparable to uncertainties originating from entropy tabular data but exceeding the uncertainties in Planck measurement. 
This discrepancy arises from the asymmetry in the scattering term $\mathcal{S_{\chi\phi}}$, which is spontaneously induced for a nonzero $c_s$.

To summarize, while our full calculation is CPU intensive, increasing the time and energy requirements,
it is worthwhile to compute the relic density of DM using a complete set of coupled Boltzmann equations for $\chi$ and $\phi$, especially in scenarios where early decoupling occurs. 
When comparing our numerical results obtained from the \texttt{Full BEs} approach with those from the \texttt{Only-$Y_{\chi}$ BEs} and \texttt{$Y_{\chi}$-and-$y_{\chi}$ BEs} approaches, we observe significant differences between the \texttt{Full BEs} and \texttt{Only-$Y_{\chi}$ BEs} methods, particularly for the forbidden DM and the resonance DM scenarios. 
However, in the case of resonance DM, the results obtained from the \texttt{Full BEs} approach are nearly identical to those obtained from the \texttt{$Y_{\chi}$-and-$y_{\chi}$ BEs} approach. 
While the differences in the secluded DM scenario are not as pronounced as in the other two scenarios, precise relic density computation still demonstrates a considerable improvement compared to standard computation methods.
We hope that future works based on this approach will increase CPU efficiencies and identify cases where a full treatment is appropriate.

\section*{Acknowledgments}
X.-C.~Duan and Y.-L.~S.~Tsai are supported by the National Key Research and Development Program of China
(No.~2022YFF0503304), and the Project for Young Scientists in Basic Research of the Chinese Academy of Sciences (No.~YSBR-092).
R.~Ramos work is supported in part by a KIAS Individual Grant (QP094601) via the Quantum Universe Center at Korea Institute for Advanced Study and by the Ministry of Science and Technology of Taiwan (Grant No.~109-2811-M-001-595).

\appendix
\section{The collision terms}
\label{sec:coll}

In this section, we present the collision terms in Eq.~\eqref{eq:boltzmann}. 

\subsection{Collision terms of $\chi$}
For the DM $\chi$, the dominant processes include annihilation into $\varphi_{\rm SM}\Bar{\varphi}_{\rm SM}$ or $\phi\phi$ 
as well as the inverse mediator decay $\chi \chi \to \phi$. 
Since the elastic scattering terms do not change the number density of $\chi$ or $\phi$, 
they can be omitted for density evolution but they have to be included for temperature evolution. 
Thus the collision term is written as
\begin{align}
C_\chi = C_\chi^{\mathrm{ann}}  + C_\chi^{\mathrm{dec}} + C_\chi^{\mathrm{el}}. \nonumber
\end{align}
In this work, the squared invariant amplitude $|\mathcal{M}|^2$ 
is summed over all internal degrees of freedom of all initial and final state particles involved.

\begin{itemize}
\item $\chi\chi$ annihilation term ($C_\chi^{\mathrm{ann}}$):\\
DM can annihilate to a pair of SM particles $\varphi$ or the new scalars $\phi$. 
Hence, the annihilation term can be written as   
\begin{align}
	C_\chi^{\mathrm{ann}}
	= {} & C_{\chi \chi \to \varphi \Bar{\varphi} } + C_{\chi \chi \to \phi \phi},
\end{align}
where
\begin{align}
	C_{\chi \chi \to \varphi \Bar{\varphi} } 
	= {} &\frac {1} {2 g_\chi}
	 \int d\Pi_\chi
	 \int d\Pi_\varphi
	 \int d\Pi_{\tilde{\varphi}} \times 
	(2 \pi)^{4} \delta^{(4)} (p_\chi + \tilde{p}_\chi - k - \tilde{k})
	\nonumber\\
	&\times
	 \left[
		|\mathcal{M}_{\chi \chi \leftarrow \varphi \Bar{\varphi}}|^2
		f_\varphi(\omega) f_{\Bar{\varphi}}(\tilde{\omega})
		- |\mathcal{M}_{\chi \chi \to \varphi \Bar{\varphi}}|^2
		f_\chi(T_\chi,E_\chi) f_\chi(T_\chi,\tilde{E}_\chi)
	\right],
	\nonumber\\
	C_{\chi \chi \to \phi \phi}
	= {} & \frac {1} {2 g_\chi}
	 \int d\Pi_\chi
	 \int d\Pi_\phi
	 \int d\Pi_{\tilde{\phi}}
	\times (2\pi)^{4} \delta^{(4)} (p_\chi + \tilde{p}_\chi - p_\phi - \tilde{p}_\phi)
	\nonumber\\
	& \times
	\left[
		|\mathcal{M}_{\chi \chi \leftarrow \phi \phi}|^2
		f_\phi(T_\phi,E_\phi) f_\phi(T_\phi,\tilde{E}_\phi)
		- |\mathcal{M}_{\chi \chi \to \phi \phi}|^2
		f_\chi(T_\chi, E_\chi) f_\chi(T_\chi,\tilde{E}_\chi)
	\right].
	\label{eq:cxann}
\end{align}
In the absence of Boson condensation or Fermi degeneracy, 
we have ignored Pauli blocking and Bose enhancement factors,
i.e., $(1\pm f_i) \approx 1$.
Here, we define 
\begin{equation}
\label{eq:dPi}
d\Pi_i\equiv \frac {d^3 \mathbf{p}_i} { (2\pi)^3 2 E_i}.
\end{equation}

\item Decay term ($C_\chi^{\mathrm{dec}}$):\\
The decay term for the WIMP $\phi\to\chi\chi$ is given by
\begin{align}
	C_\chi^{\mathrm{dec}}
	= {} & \frac {1} {2 g_\chi}
	\int d\Pi_{\tilde{\chi}} \int d\Pi_{\phi}
	(2 \pi)^{4} \delta^{(4)} (p_\phi - p_\chi - \tilde{p}_\chi)
	\nonumber\\
	& \times
	\left[
		|\mathcal{M}_{\phi \to \chi \chi}|^2 
		f_\phi(T_\phi,E_\phi) -
		|\mathcal{M}_{\phi \leftarrow \chi \chi}|^2
		f_\chi(T_\chi,E_\chi) f_\chi(T_\chi,\tilde{E}_\chi)
	\right].
\end{align}

\item Elastic scattering term ($C_\chi^{\mathrm{el}}$):\\
Since the scattering $\chi\phi\to\chi\phi$ and $\chi \varphi \to \chi \varphi$ 
do not change the number particles, this term is only used in the second-moment (temperature evolution).  
The scattering term $C_\chi^{\mathrm{el}}$ contains two components, 
$C_{\chi \varphi \to \chi \varphi}$ and $C_{\chi \phi \to \chi \phi}$. 
They can be written as 
\begin{align}
	\label{eq:xSMscatter}
	C_{\chi \varphi \to \chi \varphi}= {} & \frac {1} {2 g_\chi}
	\int d\Pi_{\tilde{\chi}} \int d\Pi_{\varphi} \int d\Pi_{\varphi'}
	(2 \pi)^{4} \delta^{(4)} (p_\chi + k - \tilde{p_\chi} - \tilde{k})
	\nonumber\\
	& \times
	|\mathcal{M}_{\chi \varphi \to \chi \varphi}|^2 
	\left[
		f_{\varphi}(\tilde{\omega}) f_\chi(T_\chi,\tilde{E_\chi}) -
		f_{\varphi}(\omega) f_\chi(T_\chi,E_\chi)
	\right], \\
	C_{\chi \phi \to \chi \phi}= {} & \frac {1} {2 g_\chi}
    	\int d\Pi_{\tilde{\chi}} \int d\Pi_{\phi} \int d\Pi_{\tilde{\phi}}
	(2 \pi)^{4} \delta^{(4)} (p_\chi + p_\phi - \tilde{p}_\chi - \tilde{p}_\phi)
	\nonumber\\
	& \times
	|\mathcal{M}_{\chi \phi \to \chi \phi}|^2 
	\left[
		f_\chi(T_\chi,\tilde{E_\chi})f_\phi(T_\phi,\tilde{E_\phi}) -
		f_\chi(T_\chi,E_\chi)f_\phi(T_\phi,E_\phi)
	\right].
	\label{eq:xmedscatter}
\end{align}
Again, we have ignored the effects of Fermi blocking and Bose enhancement factors.
\end{itemize}

\subsection{Collision terms of $\phi$}
In addition to annihilation, decay, and elastic scattering terms, 
the collision terms of the new scalar $C_\phi$ contain an co-annihilation term as well
\begin{align}
C_\phi = C_\phi^{\mathrm{ann}}  + C_\phi^{\mathrm{dec}} + C_\phi^{\mathrm{co}} + C_\phi^{\mathrm{el}}. \nonumber
\end{align}
Their explicit forms are presented below.

\begin{itemize}
\item $\phi\phi$ annihilation term ($C_\phi^{\mathrm{ann}}$):\\
The annihilation and scattering terms are analogous to $C_\chi$. 
We can simply make $\chi\leftrightarrow\phi$ in $C_{\chi}^{{\rm ann}}$ and Eq.~\eqref{eq:cxann}.

\item Decay term ($C_\phi^{\mathrm{dec}}$):\\
The new scalar $\phi$ decays into SM final states as well as WIMP pairs. 
The $\phi$ decay term can be divided in two components 
\begin{align}
	C_\phi^{\mathrm{dec}}
	= {} & C_{\phi \to \varphi \Bar{\varphi} } + C_{\phi \to \chi \chi},
\end{align}
where   
\begin{align}
	C_{\phi \to \varphi \Bar{\varphi} }
	= \frac {1} {2 g_\phi}
	& 
	\int d\Pi_\varphi
	\int d\Pi_{\tilde{\varphi}}	\times 
	(2 \pi)^{4} \delta^{(4)} (p_\phi - k - \tilde{k})
	\nonumber\\
	& \times
	\left[
		- |\mathcal{M}_{\phi \to \varphi\bar{\varphi}}|^2
		f_\phi(T_\phi,E_\phi)
		+ |\mathcal{M}_{\phi \leftarrow \varphi \bar{\varphi} }|^2
		f_{\varphi}(\omega) f_{\tilde{\varphi}}(\tilde{\omega})
	\right],\\
	C_{\phi \to \chi \chi }= \frac {1} {2 g_\phi}
	& 
	\int d\Pi_{\tilde{\chi}} \int d\Pi_{\chi}\times
	(2 \pi)^{4} \delta^{(4)} (p_\phi - p_\chi - \tilde{p}_\chi)
	\nonumber\\
	& \times
	\left[
		- |\mathcal{M}_{\phi \to \chi \chi}|^2
		f_\phi(T_\phi,E_\phi)
		+ |\mathcal{M}_{\phi \leftarrow \chi \chi}|^2
		f_\chi(T_\chi,E_\chi) f_\chi(T_\chi,\tilde{E}_\chi)
	\right].
\end{align}

\item SM-$\phi$ co-annihilation term ($C_\phi^{\mathrm{co}}$):\\
The interaction rate between $\phi$ and the SM sector can be dominated by co-annihilation processes 
involving photons and gluons~\cite{Matsumoto:2018acr}. 
The new scalar can absorb a massless gauge boson $\varphi_b$ (photon or gluon) and then 
emit a pair of SM fermions as ${\phi \varphi_b \to \varphi_f \varphi_f}$.  
Additionally, the co-annihilation with a SM fermion 
${\phi \varphi_f \to \varphi_b \varphi_f}$ can take place. 
Hence, the co-annihilation term can be written as
\begin{align}
        C_{\phi \varphi_2 \to \varphi_3 \varphi_4}
		= {} & \frac {1} {2 g_\phi}
        \int d\Pi_{\varphi_2}
        \int d\Pi_{\tilde{\varphi}_3}
        \int d\Pi_{\varphi_4}
        (2 \pi)^{4} \delta^{(4)} (p_\phi + k_2 - k_3 - k_4)
        \nonumber\\
        \times
        & \left[
                |\mathcal{M}_{\phi \varphi_2 \leftarrow \varphi_3 \varphi_4}|^2
                f_{\tilde{\varphi}_3}(\omega_{\tilde{\varphi}_3}) f_{\varphi_4}(\omega_{\varphi_4})
                - |\mathcal{M}_{\phi \varphi_2 \to \varphi_3 \varphi_4}|^2
                f_\phi(T_\phi,E_\phi) f_{{\varphi}_2}(\omega_2)
        \right].
\end{align}

\item Elastic scattering term ($C_\phi^{\mathrm{el}}$):\\
The elastic scattering does not change the number density but contributes to temperature evolution. 
Therefore, this term is only included in second-moment equations. 
The scattering term $C_\phi^{\mathrm{el}}$ 
can obtained by doing the exchange $\chi\leftrightarrow\phi$ in
Eqs.~\eqref{eq:xSMscatter} and \eqref{eq:xmedscatter}.

\end{itemize}


\section{Number density and temperature evolution equations}
\label{sec:0th_and_2nd}

The number density and temperature evolution equations can be obtained 
by integrating Eq.~\eqref{eq:boltzmann} with $\mathbf{p}_i^0$ and $\mathbf{p}_i^2/E_i$ 
over the phase space, respectively. 
In this section, we demonstrate the derivation of 
the number density evolution for $Y_\chi$ Eq.~\eqref{numberBEs} and $Y_\phi$ Eq.~\eqref{numberBEs2}. 
The temperature evolution for $T_\chi$ Eq.~\eqref{tempBEs} and $T_\phi$ Eq.~\eqref{tempBEs} are also presented.

\subsection{Number density evolution}

The number density evolution (zeroth-moment) equation is found by integrating Eq.~\eqref{eq:boltzmann} over the phase space
\begin{equation}
	g_i \int
	d\Pi_{i}\, E \left(
        \partial_t - H \mathbf{p} \cdot \nabla_\mathbf{p}
        \right) f_i = g_i \int
	d\Pi_{i}\, C_i[f_i]
	\label{eq:0th}
\end{equation}
The left hand side (LHS) of Eq.~\eqref{eq:0th} becomes 
\begin{align}
	 g_i \int \frac {d^3 \mathbf{p}} {(2\pi)^3}
	\left(
		\partial_t - H \mathbf{p} \cdot \nabla_{\mathbf{p}}
	\right) f_i(\mathbf{p})
	= 
	\frac {d n_i} {dt} + 3H n_i
	= x \tilde{H} s Y_i',
\end{align}
where we have used integration by parts and conservation of entropy, i.e., $ d(sa^3)/dt = 0 $
with $s = (2 \pi^2 / 45) g_\mathrm{eff}^s T^3 $ being the entropy density, $g_\mathrm{eff}^s$ the effective entropy degrees of freedom, and $a$ the scale factor of the universe.
The factor $x \tilde{H}$ arises from differentiation with respect to the variable $x \equiv m_\chi / T$, where we have defined
\begin{align}
    \tilde{H} \equiv H
    \left[1 + \frac{1}{3}
        \frac{d \log(g_\mathrm{eff}^s)}{d \log(T)}
    \right]^{-1}.
\end{align}
For the right hand side (RHS) of Eq.~\eqref{eq:0th}, excepting the elastic scattering term that
does not change number density, we present the 
integration of the annihilation and decay terms in what follows.


\subsubsection{The annihilation terms}

Here we summarize the annihilation terms of $\chi$ and $\phi$ in the RHS of Eq.~\eqref{eq:0th}. 

\begin{itemize}

\item $\chi \chi \to \varphi \bar{\varphi}$ and $\chi \chi \to \phi \bar{\phi}$: \\
By using Eq.~\eqref{eq:ansatz}, 
the DM annihilation to a SM pair can be rewritten as
\begin{align}
    2 g_\chi \int d\Pi_{\chi}\times
   \left( C_{\chi \chi \rightarrow \varphi \Bar{\varphi} } + C_{\chi \chi \rightarrow \phi \phi} \right)
   = {} & \avg{
		\sigma v_{\varphi \Bar{\varphi} \to \chi \chi }
	}_{T}
	 n_{\varphi, \mathrm{eq}}^2(T)
	- \avg{
		\sigma v_{\chi \chi \rightarrow \varphi \Bar{\varphi}}
	}_{T_\chi}
	 n_{\chi}^2(T_\chi) 
\nonumber\\
- & \avg{\sigma v_{\chi \chi \rightarrow \phi \phi} }_{T_\chi}
	 n_{\chi}^2(T_\chi)  +
	\avg{\sigma v_{\phi \phi\to \chi \chi}}_{T_\phi}
	 n_\phi^2(T_\phi) .  
	 \label{eq:BE_caxx}
\end{align}
The collision terms $C_{\chi \chi \to \varphi \bar{\varphi}}$ and $C_{\chi \chi \to \phi \bar{\phi}}$ 
are given in Eq.~\eqref{eq:cxann} and 
$\avg{\sigma v}_{T_i}$ is defined in Eq.~\eqref{eq:sigmav0}. 
Here, we assume $f_i$ to have Maxwellian form before kinetic decoupling.

Replacing the number density $n_i$ to the comoving number density $Y_i$ with Eq.~\eqref{eq:Yy}, 
we can obtain the first two rows of Eq.~\eqref{numberBEs}.

\item $\phi \phi \to \varphi \bar{\varphi}$ and $\phi \phi \to \chi \bar{\chi}$: \\
Similarly, the annihilation of $\phi$ can be obtained by exchanging $\chi$ and $\phi$ 
in Eq.~\eqref{eq:BE_caxx},  
\begin{align}
	2  g_\phi \int d\Pi_{\phi}
	\left(
		C_{\phi \phi \rightarrow \varphi \Bar{\varphi}}
		+ C_{\phi \phi \rightarrow \chi \bar{\chi}}
	\right)
	= {} & \avg{
		\sigma v_{\varphi \Bar{\varphi}\to \phi \phi }
		}_T n_{\varphi, \mathrm{eq}}^2(T)
	- \avg{
		\sigma v_{\phi \phi \rightarrow \varphi \Bar{\varphi}}
		}_{T_\phi} n_{\phi}^2(T_\phi) \nonumber\\
	& + \avg{
		\sigma v_{\chi \chi \rightarrow \phi \phi}
		}_{T_\chi} n_\chi^2(T_\chi)
	- \avg{
		\sigma v_{\phi \phi \rightarrow \chi \chi}
		}_{T_\phi} n_{\phi}^2(T_\phi).
	\label{eq:BE_capp}
\end{align}
The above equation is the first two rows of Eq.~\eqref{numberBEs2}. 
\end{itemize}

\subsubsection{The decay terms}
The zeroth moment of the decay term in the Eq.~\eqref{numberBEs} is 
\begin{align}
	2 g_\chi \int d\Pi_{\chi}\times
	C^{\mathrm{dec}}_\chi
	= {} & \int \frac {d^3 \mathbf{p}_\phi} { (2\pi)^3 2E_\phi}
	\int \frac {d^3 \mathbf{p}_\chi} { (2\pi)^3 2E_\chi}
	\int \frac {d^3 \tilde{\mathbf{p}}_\chi} { (2\pi)^3 2\tilde{E}_\chi}
	\times
	(2 \pi)^{4} \delta^{(4)} (p_\phi - p_\chi - \tilde{p}_\chi)
	\nonumber\\
	& \times
	|\mathcal{M}_{\phi \rightarrow \chi \chi}|^2 
	\left[
		f_\phi(T_\phi,E_\phi) -
		f_\chi(T_\chi,E_\chi) f_\chi(T_\chi,\tilde{E}_\chi)
	\right]
	\nonumber\\
	= {} & g_\phi \int \frac {d^3 \mathbf{p}_\phi} {(2\pi)^3}
		\Gamma_{\phi \rightarrow \chi \chi}\frac {n_\phi(T_\phi)} {n_{\phi, \mathrm{eq}}(T_\phi)}
		f_{\phi, \mathrm{eq}}(T_\phi,E_\phi) \nonumber \\
	 &	-g_\chi^2\int \frac {d^3 \mathbf{p}_\chi} { (2\pi)^3 }
	     \int \frac {d^3 \tilde{\mathbf{p}}_\chi} { (2\pi)^3 }
	 \sigma v_{\chi \chi \to \phi}\frac {n_\chi^2(T_\chi)} {n_{\chi, \mathrm{eq}}^2(T_\chi)}
		f_{\chi, \mathrm{eq}}(T_\chi,E_\chi)f_{\chi, \mathrm{eq}}(T_\chi,\tilde{E}_\chi)
	\nonumber \\
	= {} & 
    \avg{
		\Gamma_{\phi \rightarrow \chi \chi}
	}_{T_\phi}
	n_{\phi}(T_\phi) 
	- 
	 \avg{
	\sigma v_{\chi \chi \to \phi}
	}_{T_\chi}  n_{\chi}^2(T_\chi) 
	\label{eq:cphixx}
\end{align}
where the thermally averaged partial decay width is given by
\begin{align}
	\avg{
		\Gamma _{\phi \rightarrow \chi \chi}
	}_{T_\phi}
	& \equiv 
	\frac {g_\phi} {n_{\phi, \mathrm{eq}} (T)}
	\int \frac {d^3 \mathbf{p}_\phi} { (2\pi)^3}
	\Gamma _{\phi \rightarrow \chi \chi}
	f_\phi(T_\phi,E_\phi).
\end{align}

For the new scalar $\phi$, one can obtain the decay term in Eq.~\eqref{numberBEs2},  
by simply swapping $\chi$ and $\phi$ in Eq.~\eqref{eq:cphixx} 
but we note that the signs of two terms are different as seen in Eq.~\eqref{numberBEs2} 
for increasing and decreasing particles. 
However, the additional terms for decay into a pair of the SM final states are 
\begin{align}
	2 g_\phi \int d\Pi_{\phi}\times
	C_{\phi\to\varphi\Bar{\varphi}}
	= {}& \int \frac {d^3 \mathbf{p}_\phi} { (2\pi)^3 2E_\phi}
	\int \frac {d^3 \mathbf{k}} { (2\pi)^3 2\omega}
	\int \frac {d^3 \tilde{\mathbf{k}}} { (2\pi)^3 2\tilde{\omega}}
	\nonumber\\
	& \times
	(2 \pi)^{4} \delta^{(4)} (p_\phi - k - \tilde{k})
	|\mathcal{M}_{\phi \to \varphi \Bar{\varphi} }|^2 
	\left[
		- f_\phi(E_\phi)
		+ f(\omega) f(\tilde{\omega})
	\right]
	\nonumber\\
	= {}& - \avg{
		\Gamma_{\phi \rightarrow \varphi\Bar{\varphi}  }
	}_{T_\phi}
	n_{\phi} (T_\phi)
	+ \avg{
	\sigma v_{\varphi \varphi \to \phi}
	}_{T}  n_{\varphi, \mathrm{eq}}^2(T) .
\end{align}

\subsubsection{Co-annihilation}

The zeroth moment of $C_{\phi \varphi_2 \to \varphi_3 \varphi_4}$ term is obtained 
by a similar procedure as in the annihilation terms
\begin{align}
        & 2 g_\phi \int 
        d\Pi_\phi \times C_{\phi \varphi_2 \to \varphi_3 \varphi_4}
        \nonumber\\
        = {} & g_\phi g_{\varphi_2}
        \int \frac {d^3 \mathbf{p}_\phi} {(2\pi)^3}
        \int \frac {d^3 \mathbf{k}_{2}} { (2\pi)^3}
        \sigma v_{\phi \varphi_2 \to \varphi_3 \varphi_4}
        \left[
                f_{\varphi_3}(\omega_3)
                f_{\varphi_4}(\omega_4) -
                f_{\phi} (T_\phi, E_\phi)
                f_{\varphi_2} (\omega_2)
        \right]
        \nonumber\\
        = {} & \avg{
                \sigma v_{\varphi_3\varphi_4\to \phi \varphi_2}
        }_{T}\ 
        n_{\varphi_3, \mathrm{eq}} (T)
        n_{\varphi_4, \mathrm{eq}} (T)
        - \avg{
                \sigma v_{\phi \varphi_2 \to \varphi_3\varphi_4}
        }_{(T_\phi, T)}\ 
        n_\phi (T_\phi)
        n_{\varphi_2} (T).
\end{align}
Note that the temperatures $T$ and $T_\phi$ can be different. 
The energy distributions of two initial particles in Eq.~\eqref{eq:sigmav0} 
can be different as well.

\subsection{Temperature evolution}

The temperature evolution (second-moment) equations are found by integrating Eq.~\eqref{eq:boltzmann}
with the second moment over phase space
\begin{equation}
	g_i \int d\Pi_{i}
	\times
	\frac{\mathbf{p}_i^2}{E_i}
    \times
    E \left(
        \partial_t - H \mathbf{p} \cdot \nabla_\mathbf{p}
        \right) f_i
        =
	g_i \int d\Pi_{i}
	\times
	\frac{\mathbf{p}_i^2}{E_i}
    \times
    C_i[f_i]
	\label{eq:tempBE}
\end{equation}
The LHS of Eq.~\eqref{eq:tempBE} becomes  
\begin{align}
	& 3 \frac {d (T_i n_i)} {dt}
	+ 15 H T_i n_i
	- H \avg{
		\frac {\mathbf{p}_i^4} {E_i^3}
	}
	n_i\nonumber\\
	= {}& 3 n_i T_i \left(
		\frac {\Dot{Y}_i} {Y_i}
		+ \frac {\Dot{y}_i} {y_i}
		- 5 H
		\right)
	+ 15 H T_i n_i
	- H \avg{
		\frac {\mathbf{p}_i^4} {E_i^3}
	}
	n_i
	\nonumber\\
	= {}& 3 x \tilde{H} n_i T_i
	\left(
		\frac {Y_i'} {Y_i}
		+ \frac {y_i'} {y_i}
	\right)
	- H \avg{
		\frac {\mathbf{p}_i^4} {E_i^3}
	}
	n_i.
\end{align}
Here, we have used integration by parts and the energy-momentum relation
$E_i^2=m_i^2+\mathbf{p}_i^2$. The thermal average of $\mathbf{p}_i^4 / E_i^3$ is given by
\begin{equation}
    \avg{\frac{\mathbf{p}_i^4}{E_i^3} }
    \equiv
    \frac{g_i}{n_i}
	\int \frac {d^3 \mathbf{p}_i} {(2\pi)^3}
    \frac{\mathbf{p}_i^4}{E_i^3}
    f_i(E_i).
\end{equation}

In the following, we show the RHS of Eq.~\eqref{eq:tempBE}. 
All the collision terms, including elastic scattering, will contribute to the second-moment equations.

\subsubsection{Elastic scattering}

Considering a process 
$\chi(\mathbf{p}_\chi)+\phi(\mathbf{p}_\phi)\to 
\chi^\prime(\Tilde{\mathbf{p}}_\chi)+\phi^\prime(\Tilde{\mathbf{p}}_\phi)$, 
the temperature term $\mathbf{p}_\chi^2/E_\chi$ can be easily obtained while  
$\Tilde{\mathbf{p}}_\chi^2/\Tilde{E}_\chi$ requires some additional algebra work as shown below.
The integrals over $\Tilde{\Pi}_\chi$ and $\Tilde{\Pi}_\phi$ of 
Eq.~\eqref{eq:Schiphi} can also be directly computed as 
an integral over the solid angle of the outgoing $\chi$ particle 
\begin{align}
	& 2\int d\Tilde{\Pi}_\chi d\Tilde{\Pi}_\phi
	\left(
		\frac{\Tilde{\mathbf{p}}_\chi^2} {\Tilde{E}_\chi}
		- \frac{\mathbf{p}_\chi^2} {E_\chi}
	\right)
	(2\pi)^4 \delta^{(4)}
	(p_\chi + p_\phi - \Tilde{p}_\chi - \Tilde{p}_\phi)
	| \mathcal{M}_{\chi \phi \to \chi \phi} |^2
	\nonumber\\
	= {} &
	2\int 
	\frac {\Tilde{\mathbf{p}}_\chi^2 d\Omega}
    	{16\pi^2 \Tilde{E}_\chi \Tilde{E}_\phi }
	\left(
		\frac{\Tilde{\mathbf{p}}_\chi^2} {\Tilde{E}_\chi}
		- \frac{\mathbf{p}_\chi^2} {E_\chi}
	\right)
	\left|
	    \frac { |\Tilde{\mathbf{p}}_\chi| } {\Tilde{E}_\chi}
	    + \frac { |\Tilde{\mathbf{p}}_\chi|
    	    - |\mathbf{p}_\chi + \mathbf{p}_\phi| \cos\alpha}
	    {\Tilde{E}_\phi}
	\right| ^{-1}
	| \mathcal{M}_{\chi \phi \to \chi \phi} |^2
	\nonumber\\
	\label{eq:scatinteg}
	= {} &
	2\int \frac{ d\Omega} {16\pi^2}
	\frac {
        \Tilde{\mathbf{p}}_\chi^4 / \Tilde{E}_\chi
        - \mathbf{p}_\chi^2 \Tilde{\mathbf{p}}_\chi^2 / E_\chi }
    	{\left||\Tilde{\mathbf{p}}_\chi| (E_\chi + E_\phi)
    	    - |\mathbf{p}_\chi + \mathbf{p}_\phi| \Tilde{E}_\chi \cos\alpha \right| }
	| \mathcal{M}_{\chi \phi \to \chi \phi} |^2,
\end{align}
where $\alpha$ is defined as the angle between $\mathbf{p}_\chi + \mathbf{p}_\phi$ and $\Tilde{\mathbf{p}}_\chi$.

We can solve $\Tilde{\mathbf{p}}_\chi$ by choosing a coordinate system where the vector $\mathbf{p}_\chi$ is on the z-axis. 
The relevant vectors have the following characteristics
\begin{eqnarray}
\frac{\mathbf{p}_\chi}{|\mathbf{p}_\chi|} &=&  \left( 0,\,0,\,1\right), \nonumber\\
\frac{\mathbf{p}_\chi + \mathbf{p}_\phi}{|\mathbf{p}_\chi + \mathbf{p}_\phi|} &=&  
\left( \sin\beta,\,0,\,\cos\beta \right), \nonumber\\
\frac{\Tilde{\mathbf{p}}_\chi}{|\Tilde{\mathbf{p}}_\chi|} &=&  
\left( \sin\theta_{\chi'}\cos\phi_{\chi'},\,\sin\theta_{\chi'}\sin\phi_{\chi'},\,\cos\theta_{\chi'}\right),
\end{eqnarray}
where $\beta$ is the angle between $\mathbf{p}_\chi + \mathbf{p}_\phi$ and $\mathbf{p}_\chi$. 
The angular coordinates of $\Tilde{\mathbf{p}}_\chi$ in spherical coordinates 
are $\theta_{\chi'}$ and $\phi_{\chi'}$.
We also have 
\begin{align}
	\cos\alpha = {}& \cos\phi_{\chi'} \sin\theta_{\chi'} \sin\beta + \cos\theta_{\chi'} \cos\beta
	\label{eq:cosalpha}
\end{align}
and
\begin{align}
	\cos\beta = {}& \frac{ |\mathbf{p}_\chi| + |\mathbf{p}_\phi| \cos\theta_{12} }
	{ |\mathbf{p}_\chi + \mathbf{p}_\phi | }.
	\label{eq:cosbeta}
\end{align}
Using the Lorentz scalar
\begin{align}
	\Tilde{p}_\chi \cdot \left( p_\chi + p_\phi \right)
	= {}& \Tilde{E}_\chi (E_\chi + E_\phi)
	- |\Tilde{\mathbf{p}}_\chi| |\mathbf{p}_\chi + \mathbf{p}_\phi| \cos\alpha,
\label{eq:Lorentzscalar1}
\end{align}
the invariant product can be simply obtained in the center of mass frame,
\begin{align}
\Tilde{p}_\chi \cdot \left( p_\chi + p_\phi \right)
    = \frac{s + m_\chi^2 - m_\phi^2} {2}.
\label{eq:Lorentzscalar2}
\end{align}
Therefore, we can obtain $|\Tilde{\mathbf{p}}_\chi|$ by solving the quadratic equation
\begin{align}
\label{eq:pchieqn}
    (m_\chi^2 + |\Tilde{\mathbf{p}}_\chi|^2 ) (E_\chi + E_\phi)^2 =
    \left( \frac{s + m_\chi^2 - m_\phi^2} {2} + 
        |\Tilde{\mathbf{p}}_\chi| |\mathbf{p}_\chi + \mathbf{p}_\phi| \cos\alpha \right)^2
\end{align}
with $|\mathbf{p}_\chi + \mathbf{p}_\phi| = [(E_\chi + E_\phi)^2 - s]^{1/2}$
and $\cos\alpha$ as given in Eq. (\ref{eq:cosalpha}).
Note that the integral in Eq.~\eqref{eq:scatinteg} has to be evaluated
using all the valid (positive and real) $|\Tilde{\mathbf{p}}_\chi|$ solutions.
For Eq.~\eqref{eq:pchieqn} it is possible to have from none to at most two such solutions.

Other scattering terms in the temperature evolution such as $\mathcal{S}_{\chi\varphi}$, 
$\mathcal{S}_{\phi\chi}$, and $\mathcal{S}_{\phi\varphi}$ can be obtained in an analogous manner. 
In this work, we have performed these multidimensional integrals numerically using the Monte Carlo integration package \texttt{Cuba}.

\subsubsection{Annihilation}
Mimicking the case of zeroth moment, the second moment of the annihilation term in Eq.~\eqref{tempBEs} is derived as 
\begin{align}
	& 2 g_\chi \int d\Pi_{\chi}
	\frac {\mathbf{p}_\chi^2} {E_\chi}
	C^{\mathrm{ann}}_{\chi}
	\nonumber\\
	= {}& 3 \avg{
		T_\chi\sigma_{\varphi \Bar{\varphi}\to \chi \chi} v
	}_T
	n_{\varphi, \mathrm{eq}}^2 (T) 
	- 3 \avg{
		T_\chi\sigma v_{\chi \chi \rightarrow \varphi \Bar{\varphi}}
	}_{T_\chi}
	n_\chi^2 (T_\chi) 
	\nonumber\\
		 &+ 3 \avg{T_\chi
		\sigma v_{\phi \phi \rightarrow \chi \chi}
	}_{T_\phi}
	n_\phi^2 (T_\phi)
	- 3 \avg{T_\chi
		\sigma v_{\chi \chi \rightarrow \phi \phi}
	}_{T_\chi}
	n_\chi^2 (T_\chi),
\end{align}
where $T$ is defined as shown in Eq.~\eqref{eq:n_and_T}.
Similarly, the second annihilation moment of $\phi$ in Eq.~\eqref{tempBEs2} can be obtained by simply swapping $\chi$ and $\phi$.

\subsubsection{Decay}
The second moment of the decay terms for DM sector temperature can be obtained as follows:
\begin{align}
	2 g_\chi \int d\Pi_{\chi}
	\frac {\mathbf{p}_\chi^2} {E_\chi}
	C^{\mathrm{dec}}_{\chi}
	={} & 2\int \frac {d^3 \mathbf{p}_\phi} { (2\pi)^3 2E_\phi}
	\int \frac {d^3 \mathbf{p}_\chi} { (2\pi)^3 2E_\chi}
	\frac {\mathbf{p}_\chi^2} {E_\chi}
	\int \frac {d^3 \tilde{\mathbf{p}}_\chi} { (2\pi)^3 2\tilde{E}_\chi}
	\nonumber\\
	& \times
	(2 \pi)^{4} \delta^{(4)} (p_\phi - p_\chi - \tilde{p}_\chi)
	|\mathcal{M}_{\phi \rightarrow \chi \chi}|^2 
	\left[
		f_\phi(E_\phi)
		- f_\chi(E_\chi) f_\chi(\tilde{E}_\chi)
	\right]
	\nonumber\\
	={} & 
3 \avg{T_\chi
		\Gamma_{\phi \rightarrow \chi \chi}
	}_{T_\phi}
	n_{\phi}(T_\phi) 
	- 
3  \avg{T_\chi
	\sigma v_{\chi \chi \to \phi}
	}_{T_\chi}  n_{\chi}^2(T_\chi) 
	\label{eq:Tphixx}
\end{align}
where the thermally averaged product of $T_\chi$ and decay width $\Gamma$ is given by
\begin{align}
	\avg{T_\chi
		\Gamma _{\phi \rightarrow \chi \chi}
	}_{T_\phi}\equiv\frac{2}{n_{\phi, \mathrm{eq}}}
		& \int \frac {d^3 \mathbf{p}_\phi} { (2\pi)^3 2E_\phi}
	\int \frac {d^3 \mathbf{p}_\chi} { (2\pi)^3 2E_\chi}
	\frac {\mathbf{p}_\chi^2} {3E_\chi}
	\int \frac {d^3 \tilde{\mathbf{p}}_\chi} { (2\pi)^3 2\tilde{E}_\chi}
	\nonumber\\
	& \times
	(2 \pi)^{4} \delta^{(4)} (p_\phi - p_\chi - \tilde{p}_\chi)
	|\mathcal{M}_{\phi \rightarrow \chi \chi}|^2 
	f_{\phi, \mathrm{eq}}(E_\phi).
	\label{eq:avg_T_ga}
\end{align}

Similarly, for the temperature of $\phi$ sector, 
the second moment of its decay term can be obtained as follows
\begin{align}
	g_\phi & \int \frac {d^3 \mathbf{p}_\phi} {(2\pi)^3}
	\frac {\mathbf{p}_\phi^2} {E_\phi^2}
	C^{\mathrm{dec}}_{\phi}
	\nonumber\\
	= {}& -3\avg{
		T_\phi\Gamma_{\phi \rightarrow \varphi\varphi }
	}_{T_\phi}
	n_{\phi} (T_\phi) 
	+ 3  \avg{T_\phi
	\sigma v_{\varphi \varphi \to \phi}
	}_{T}  n_{\varphi,\mathrm{eq}}^2(T)
	\nonumber\\
	  & 	
-3 \avg{T_\phi
		\Gamma_{\phi \rightarrow \chi \chi}
	}_{T_\phi}
	n_{\phi}(T_\phi) 
	+ 
3  \avg{T_\phi
	\sigma v_{\chi \chi \to \phi}
	}_{T_\chi}  n_{\chi}^2(T_\chi). 	
	\label{eq:T_phi_to_all}
\end{align}
The thermally averaged second moment of the mediator decay width $\Gamma_\phi$ is 
similar to Eq.~\eqref{eq:avg_T_ga}.

\subsubsection{Co-annihilation}

The second moment of the $\phi$ and SM $\varphi$ scattering term is derived as follows:
\begin{align}
	& 2 g_\phi \int d\Pi_{\phi}
	\frac {\mathbf{p}_\phi^2} {E_\phi}
	C_{\phi\varphi_2 \to \varphi_3 \varphi_4}
	\nonumber\\
	= {}& g_\phi g_{\varphi_2}
	\int \frac {d^3 \mathbf{p}_\phi} {(2\pi)^3}
	\frac {\mathbf{p}_\phi^2} {E_\phi}
	\int \frac {d^3 \mathbf{k}_2} { (2\pi)^3}
\sigma v_{\phi \varphi_2 \to \varphi_3 \varphi_4}
        \left[
                f_{\varphi_3}(\omega_3)
                f_{\varphi_4}(\omega_4) -
                f_{\phi} (T_\phi, E_\phi)
                f_{\varphi_2} (\omega_2)
        \right]
	\nonumber\\
	={} & 3\avg{T_\phi\sigma v_{\varphi_3\varphi_4\to \phi \varphi_2} }_{T}\ 
        n_{\varphi_3, \mathrm{eq}} (T)
        n_{\varphi_4, \mathrm{eq}} (T)
        - 3\avg{T_\phi\sigma v_{\phi \varphi_2 \to \varphi_3\varphi_4}
        }_{(T_\phi, T)}\ 
        n_\phi (T_\phi)
        n_{\varphi_2} (T).
\end{align}

\section{The temperature tables of $\chi$-$\phi$ elastic scattering}
\label{sec:scattering}

\begin{figure}
    \centering
    \includegraphics[scale=0.40]{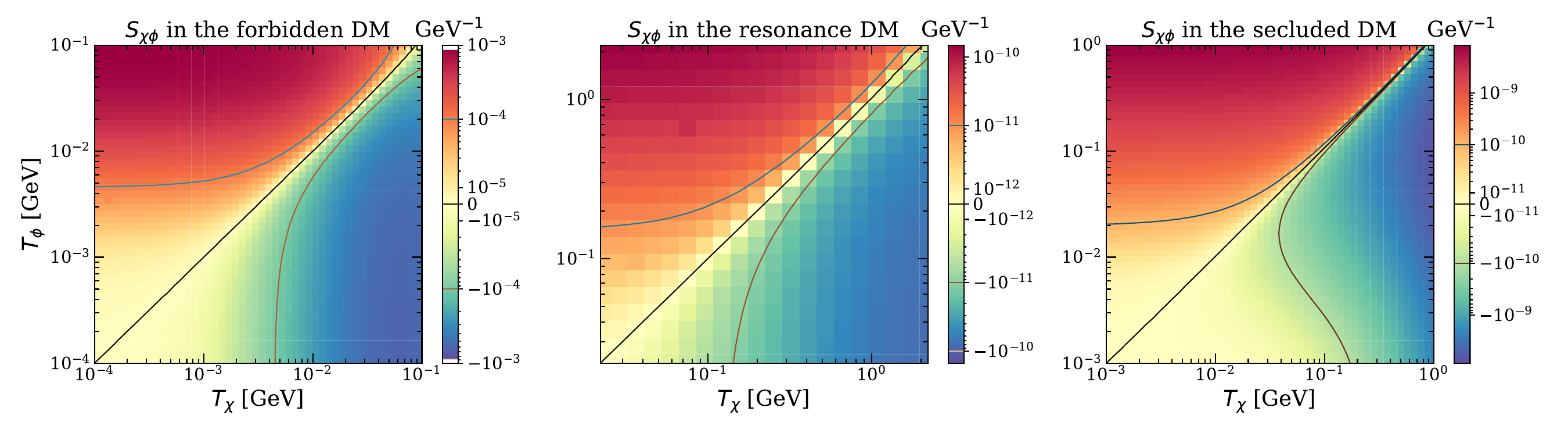}
    \caption{The resulting numerical maps for $\mathcal{S_{\chi\phi}}$
    for the forbidden scenario ($m_\chi = 0.1$~GeV, 
    $m_\phi = 0.13$~GeV, $\sin\theta = 10^{-3}$, $c_s = 0.1$ and $\lambda_{\phi H}=1.0$ ) in the left panel, 
    the resonance scenario ($m_\chi = 2.2\gev$,
    $m_\phi = 4.7\gev$, $\sin\theta = 0.01$, $c_s = 10^{-3}$ and $\lambda_{\phi H}=1.0$ ) in the center panel 
    and the secluded scenario ($m_\chi = 1.0$\gev, $m_\phi = 0.01$\gev,  $\sin\theta = 10^{-9}$, $c_s = 0.045$ and $\lambda_{\phi H} = 0.1$) in the right panel.
    The orange, black, and blue lines correspond to the contours 
    where $\mathcal{S_{\chi\phi}} = \{-10^{-4},~0,~10^{-4}\}\gev^{-1}$ in the forbidden scenario, $\mathcal{S_{\chi\phi}} = \{-10^{-11},~0,~10^{-11}\}\gev^{-1}$ in the resonance scenario
    and $\mathcal{S_{\chi\phi}} = \{-10^{-10},~0,~10^{-10}\}\gev^{-1}$ in the secluded scenario.
    }
    \label{fig:schiphi}
\end{figure}

In Fig.~\ref{fig:schiphi}, we present $\mathcal{S}_{\chi\phi}$ 
on the plane ($T_\chi$, $T_\phi$) for three benchmark scenarios,  
$m_\chi\approx m_\phi$ (left panel), resonance $m_\phi\approx 2 m_\chi$ (center panel) and secluded $m_{\phi}\ll m_{\chi}$ (right panel). 
From Eq.~\eqref{eq:Schiphi}, we learned that DM gains energy from $\phi$ if $\mathcal{S}_{\chi\phi}>0$. 
On the other hand, energy is transferred from dark sector to $\phi$ sector 
if $\mathcal{S}_{\chi\phi}<0$. 
As expected from the second law of thermodynamics,
we can see that the energy always flows from a sector with higher temperature 
to other with lower temperature resulting in $\mathcal{S}_{\chi\phi}$ symmetric around the diagonal in both the forbidden and the resonance scenarios.
In these two scenarios, DM mass and the new scalar mass are in the same order, 
while in the
secluded scenario $\phi$ is much lighter. 
Therefore, $\phi$ becomes non-relativistic later than DM 
and the resulting an asymmetric  $\mathcal{S}_{\chi\phi}$ region.

\bibliography{4BEs_v3}
\end{document}